\documentclass[12pt,preprint]{emulateapj}
\def\gsim{\;\rlap{\lower 2.5pt
 \hbox{$\sim$}}\raise 1.5pt\hbox{$>$}\;}
\def\lsim{\;\rlap{\lower 2.5pt
   \hbox{$\sim$}}\raise 1.5pt\hbox{$<$}\;}
\def\Ha{$\rm{H}\alpha$}
\def\mum{$\mu$m }

\def\Ks{K$_{\rm{s}}$}
\def\D25{$\rm{D}_{25}$}
\def\Msun{$\rm{M}_{\odot}$}
\def\sdunits{$\rm{M}_{\odot}\rm{pc}^{-2}$}
\def\as2{arcsec$^{-2}$}
\begin{document}

\title{Spiral-Induced Star Formation in the Outer Disks of Galaxies}
\author{Stephanie J. Bush\altaffilmark{1}, T.J. Cox\altaffilmark{1,2,4}, Christopher C. Hayward \altaffilmark{1}, David Thilker\altaffilmark{3}, Lars Hernquist\altaffilmark{1}, Gurtina Besla \altaffilmark{1}}
\altaffiltext{1}{Harvard-Smithsonian Center for Astrophysics, 60 Garden St, Cambridge, MA 02143 USA}
\altaffiltext{2}{W.M. Keck Postdoctoral Fellow at the Harvard-Smithsonian Center for Astrophysics}
\altaffiltext{3}{Center for Astrophysical Sciences, The Johns Hopkins
  University, 3400 North Charles ST, Baltimore, MD 21218}
\altaffiltext{4}{Currently at Carnegie Observatories, 813 Santa
  Barbara St., Pasadena, CA 91101}
\email{sbush@cfa.harvard.edu}

\slugcomment{Accepted for publication in the Astrophysical Journal}

\begin{abstract}

The outer regions of galactic disks have received increased attention
since ultraviolet observations with GALEX demonstrated that nearly
30\% of galaxies have UV emission beyond their optical extents,  
indicating star formation activity.  These galaxies have been
termed extended UV (XUV) disks.  Here, we address whether these
observations contradict the gas surface density threshold for star
formation inferred from \Ha \ radial profiles of galaxies.  We run smoothed particle
hydrodynamics simulations of isolated disk galaxies with fiducial star
formation prescriptions and show
that over-densities owing to the presence of spiral structure can
induce star formation in extended gas disks. For direct comparison with observations, we use the 3-D
radiative transfer code \textsc{Sunrise} to create simulated FUV and
\Ks -band images.  We find that galaxies classified as Type I XUV
disks are a natural consequence of spiral patterns, but we are unable
to reproduce Type II XUV disks.  We also compare our results to
studies of the Kennicutt-Schmidt relation in outer disks.

\end{abstract}


\section{Introduction}

Star formation rate profiles derived from \Ha \ emission
indicate that galaxies follow a ``Kennicutt-Schmidt Law'' where the
star formation rate and gas surface densities averaged in
azimuthal annuli are related by
$\Sigma_{\rm{SFR}} \propto \Sigma_{\rm{gas}}^{1.5}$ \citep{Kennicutt-1998, Kennicutt-1989}. However, sudden drops in \Ha \
intensity create strong departures
at a gas surface density of 3-5 \sdunits. This is generally interpreted as a threshold density for star
formation \citep{Kennicutt-1989, Martin-Kennicutt-2001}. Several
explanations for this threshold have been proposed, including a transition between dynamically
unstable and stable regions of the galaxy \citep[e.g.][]{Toomre-1964}
or a phase transition of the gas
\citep[e.g.][]{Elmegreen-Parravano-1994, Schaye-2004, Krumholz-et-al-2009}.

Outer disks have consequently been considered inhospitable environments for
star formation despite the existence of \Ha \ knots at large radii
\citep{Kennicutt-1989,Martin-Kennicutt-2001, Ferguson-et-al-1998}.
Interest in outer disks was rekindled by observations with the Galaxy
Evolution Explorer (GALEX: \citet{Martin-et-al-2005}) which revealed
that $\sim 30$\% of disk galaxies have UV emitting sources beyond their
optical disks \citep{Thilker-et-al-2007, Gildepaz-et-al-2005,
Thilker-et-al-2005, Zaritsky-Christlein-2007}.  In the case of M83 and
NGC 4625, UV knots have been identified as low mass stellar complexes
and, if visible in the \Ha, are generally ionized by a single star \citep{Gildepaz-et-al-2007}.
These knots are dynamically cold and
rotating, indicating that outer disk complexes are extensions of the inner disk
\citep{Christlein-Zaritsky-2008}.

\citet[][hereafter T07]{Thilker-et-al-2007} classified extended UV
(XUV) disks into two types. In 2/3 of XUV disks, the UV emission is
structured, lying in filamentary, spiral patterns (Type I XUV
disks). In the other 1/3, a large zone  in the outer regions of the galaxy is
dominated by UV emission (Type II
XUV disks).

The prevalence of star formation in outer disks raises new questions
about the nature of star formation in diffuse environments. In particular, we address whether a threshold density to star formation is consistent with the observed emission. \citet[][hereafter B07]{Boissier-et-al-2007} selected a sample of
galaxies of large angular size, including XUV and non-XUV disks, and
studied their extinction corrected UV and gas profiles. While the uncertainties are large, B07 do not
find a departure from the Kennicutt-Schmidt relation at low gas density. They suggest
that the high mass end
of the stellar initial mass function (IMF) is
not sampled in every stellar cluster at the low star formation rates observed and \Ha \ emitting
stars are not formed in all star forming complexes. They argue that this is responsible for the
difference in the UV and \Ha \ profiles and the
Kennicutt-Schmidt law is followed to low surface density. 

\citet{Pflamm-Kroupa-2008}
propose instead that the Kennicutt-Schmidt law is obeyed below the
traditional threshold density but that high mass stars capable of
emitting in the \Ha \ simply do not form at low gas
densities. This is supported by a deficiency of H$\alpha$
emission in dwarf galaxies that cannot be explained by IMF
sampling effects \citep{Lee-et-al-2009}. \citet{Roskar-et-al-2008b} note that stars
migrate radially from their formation sites in the disk, which can
modify the stellar populations of outer disks. However, it is not
clear that this process is efficient enough to relocate UV emitting
stars to 2-3 \D25 \citep{Roskar-et-al-2008a, Roskar-et-al-2008b}.

Alternatively, at some sites the gas density
may exceed a star formation threshold locally, allowing
stars to form beyond the radius where the azimuthally averaged gas
density is at or below a threshold density \citep{Kennicutt-1989,
Martin-Kennicutt-2001, Schaye-2004, Elmegreen-Hunter-2006,
Gildepaz-et-al-2007}. Outer disk stellar complexes are often
coincident with local H\,I over-densities
\citep[][T07]{Ferguson-et-al-1998}, supporting this interpretation. XUV disks are gas rich for their star formation
rates, indicating that a high fraction of their gas is not
undergoing star formation (T07). 
Studies of UV selected sources in the XUV disk of M\,83 indicate that
star formation occurs only above a critical gas density defined by the
Toomre criterion for dynamical instability \citep{Dong-et-al-2008}
and, where it occurs, follows a Kennicutt-Schmidt law.

This theory has been explored by \citet{Elmegreen-Hunter-2006}.  In
their model, gas clumping triggered by spiral density waves, radial
variations in the interstellar medium (ISM) turbulence, and gas phase
transitions lead to localized regions of active star formation in
extended gas disks.  They implement a Kennicutt-Schmidt law and a radially varying threshold density based on the
Toomre criterion \citep{Elmegreen-Hunter-2006}. These processes enable star
formation to extend into the outer disk. In \citet{Bush-et-al-2008},
we used smoothed particle hydrodynamic (SPH) simulations of an
isolated disk galaxy with an extended gas disk of initially constant
surface density to explore the morphology resulting from such in situ outer disk
star formation.  We showed that spiral structure can propagate from
the inner disk to the extended gas disk, producing regions of locally
enhanced gas density and triggering star formation in filaments similar to those observed in Type I XUV disks. However, this study was restricted to
one galaxy model, limiting our ability to draw general conclusions.

In this paper, we expand on the work of \citet{Bush-et-al-2008} by
varying the structure of the simulated galaxy, in particular the gas
profile and the disk to halo mass ratio, to explore the amount and
morphology of outer disk star formation that result naturally without
altering traditional star formation laws. In addition, we produce
simulated images to directly compare UV and \Ks \ emitting populations to
observations. This allows us to classify our models as normal, Type I or II XUV disks. We also compare
our results to studies of the Kennicutt-Schmidt law, radially and
locally. While we test whether the
UV and \Ks \ emission of XUV disks can be reproduced using traditional
star formation laws, we do not create H$\alpha$ images and therefore
do not address the discrepancy between H$\alpha$ and UV radial
profiles. In \S 2 we
describe our models, in \S 3 we describe the evolution and
uncertainties in one model in detail, in \S 4 we describe the full set
of models and how they compare to observations and finally in
\S 5 and \S 6 we discuss our results. Our models succeed in producing the UV morphologies of Type I XUV disks. However, they do not reproduce the UV bright outskirts of Type II XUV disks.

\section{Method} \label{sec:method}

\subsection{SPH Simulations}

We use the code \textsc{Gadget2} \citep{Springel-2005}, based on the
fully conservative formulation of SPH developed by
\citet{Springel-Hernquist-2002} to run simulations of galaxies with
extended gas disks. We incorporate a sub-resolution multi-phase model
of the ISM and star formation according to \citet[][hereafter
SH03]{Springel-Hernquist-2003}, which includes radiative cooling as in
\citet{Katz-et-al-1996} and \citet{Dave-et-al-1999}.  Star formation
from the cold phase gas (total cold gas, our simulation does not treat
atomic and molecular gas separately) follows a Schmidt volume density
law $\rho_{\rm{SFR}} \propto \rho_{\rm{gas}}^{N}$ with $N=1.5$ and
normalized to a star formation timescale ($t_{\star}$).

We run simulations with two different star formation timescales
to explore how $t_{\star}$ affects the XUV emission. One set of
simulations adopts the values proposed in SH03: $t_{\star} = 2.1$ Gyr,
$A_{0}= 1000$, and $T_{SN} = 10^{8}$, where $A_{0}$ and $T_{SN}$ are
parameters that regulate the mass fraction in the cold phase of the
ISM.  This normalizes the star formation rate to the best-fit relation
found in \citet{Kennicutt-1998}. The second set of simulations are run
with parameters which normalize the star formation rate roughly to
that of the Milky Way (MW), since these are MW like models: $t_{\star} = 6.25$ Gyr, $A_{0}= 3000$, and
$ T_{SN} = 3\times10^{8}$.  This results in lower levels of star
formation for a given gas density.  The local star formation volume
density cutoff corresponds to $\sim 0.004$ M$_{\odot}$/pc$^{3}$ for
both sets of parameters. In all cases, star formation rates lie well
within the scatter shown in \cite{Kennicutt-1998} and the surface
density threshold falls between $\sim$ 3-5 M$_{\odot}$/pc$^{2}$,
consistent with observations \citep{Kennicutt-1989,
Martin-Kennicutt-2001}. The effective equation of state for the
star-forming gas is determined from these parameters through the
pressurizing action of supernova feedback, allowing us to stably
evolve disks with even very large gas fractions 
\citep{Springel-Hernquist-2005}.

\subsection{SPH Initial Conditions} \label{sec:ics}

Our galaxy models are constructed using the procedure outlined in
\citet{Springel-Dimatteo-Hernquist-2005}. We construct analogs to the
Milky Way with a total mass of $M_{200}\sim10^{12}$ M$_{\odot}$
composed of a \citet{Hernquist-1990} dark matter halo with
concentration $c=9$, an exponential stellar disk with scale length
3.75 kpc (we adopt the Hubble parameter $h=0.72$ throughout this paper), and an exponential
gas disk of the same scale length. We then add an extended gas disk 
either of approximately constant density or following an $r^{-1}$
\citet{Mestel-1963} radial profile that starts at a ``break radius'' of
3.5 scale lengths ($\sim 13$ kpc) and extends to 8.7 scale lengths
($\sim$ 32 kpc). We also create a control model with no extension
(pure exponential profile). We then normalize with each of
these outer disk profiles (pure exponential, flat, and Mestel) to have
a gas surface density of 3, 5 and 10 \sdunits at the break radius,
spanning the range of outer disk profiles seen in the THINGS
\citep{Walter-et-al-2008} and \citet{Broeils-vanWoerden-1994}
surveys. Table~\ref{tab:ics} summarizes these properties and lists a
run-name for each model. We do not simulate a galaxy with a high mass
constant density extension because nearly constant density disks are
not found at those densities.  The gas content and distribution agree
roughly with the $M_{\rm{HI}}/M_{\rm{tot}}$ and $\log M_{\rm{HI}}/\log
D_{\rm{HI}}$ trends seen in \citet{Broeils-Rhee-1997}. The flat-low
model is very similar to the model shown in \citet{Bush-et-al-2008}.

The ratio of the mass of the inner gas disk to the mass of the inner
stellar disk is fixed at $\sim$ 1:5, meaning the overall stellar disk mass is increased
when the gas surface density is increased. The halo mass is also
increased accordingly to keep the halo to disk mass ratio at
constant at 0.041. Varying the halo to disk mass ratio changes the
spiral structure amplitude, which we do in \S~\ref{sec:spiralstruct}.  A stellar
\citet{Hernquist-1990} bulge with a scale length of 0.27 kpc is added
with 0.008 of the total mass of the simulation, making it $\sim25$\%
the mass of the stellar disk.  Particle numbers are varied for each
model to keep the mass per particle constant.  The mass per particle
in the stellar and gas disks is $\sim 10^{5} M_{\odot}$ and in the
halo is $\sim 10^{7} M_{\odot}$.  The \textsc{Gadget} softening length
of the baryonic particles and halo particles are 0.14 kpc and 0.28
kpc, respectively.  The models each have a 0.01 $M_{200}$ mass black
hole in its center. The black hole is represented by a sink particle
that can accrete mass at the Bondi rate
\citep{Springel-Dimatteo-Hernquist-2005, Dimatteo-et-al-2005}, but as
we are interested in the outer disk its presence does not influence
our results.

\begin{table}
\begin{center}

\begin{tabular}{lcccc}
\hline 
run name & gas profile shape & \multicolumn{2}{c}{ surface density at break radius for } &  \\
& & gas \sdunits & stars \sdunits & \\

\hline
\hline
exp-low                                         &   exponential  &   3    &  12    \\
exp-mid                                         &   exponential  &   5    &  20    \\
exp-high                                        &   exponential  &   10   &  40    \\
\hline
Mestel-low                                      &   Mestel     &   3     &   12    \\
Mestel-med                                      &   Mestel     &   5     &   20    \\
Mestel-high                                     &   Mestel     &   10    &   40    \\
\hline
flat-low                                        &   flat       &   3     &   12    \\
flat-mid                                        &   flat       &   5     &   20    \\
\hline 
\end{tabular}
\end{center}
\caption{Gas distributions for models used in this study. Every model is initialized with an exponential inner disk profile out to 13 kpc and the outer disk profile listed here to 32 kpc. The mass of each component of the galaxy is scaled to give the surface density listed here at the break radius and the number of particles in each component is scaled to keep the mass per particle constant in each component. Details are described in \S~\ref{sec:ics}. All models are evolved in isolation.}

\label{tab:ics}

\end{table}

\subsection{Radiative Transfer}

To create simulated FUV and \Ks -band images, we employ the 3D
adaptive grid polychromatic Monte Carlo radiative transfer code
\textsc{Sunrise}\footnote{Sunrise is a free, publicly available software
that can be applied to any hydrodynamic galaxy simulation.
http://sunrise.familjenjonsson.org.} \citep{Jonsson-2006,
Jonsson-et-al-2009}. \textsc{Sunrise} assigns a spectrum to each
particle in our simulations that would be a source of radiation
(stellar and black hole particles) and then propagates photon
``packets'' from these sources through the dusty ISM, represented by the gas
particles, using a Monte Carlo approach.  The black hole particle
emits an empirical template AGN spectrum derived from observations of
quasars \citep{Hopkins-et-al-2007} but since we are examining the
outer disk, and this is a small black hole with almost negligible
accretion, this has negligible impact on our results.  

Each stellar particle over 10 Myr old is given a spectrum
corresponding to its age and metallicity from \textit{Starburst 99}
\citep{Leitherer-et-al-1999} population synthesis models. We assume a
Kroupa IMF \citep{Kroupa-Tout-Gilmore-1993}.  Star particles that
existed at the beginning of the simulation are assigned an age from a
distribution that is determined by a star formation history chosen to match observations of local disks (see
\citet{Rocha-et-al-2008} for more details).  The disk was assigned an
exponentially increasing star formation rate with a timescale of 106
Gyr (which makes the star formation rate effectively constant) with the
oldest stars being 13.9 Gyr old.  
The bulge was assumed to form in 
an instantaneous burst with an age of 13.9 Gyr. Since these parameters
are not given any radial gradient, and the number of stars in the
outer disk is low, they only slightly affect our results.

Star clusters with ages less than 10 Myr are assumed to be located in
their nascent birth-clouds of molecular gas and are given a modified
spectrum which accounts for the effects of H\,II and
photo-dissociation regions (PDRs). The evolution of H\,II regions and
PDRs are described analytically and the photo-ionization code
\textsc{MAPPINGS III} \citep{Groves-et-al-2004} is used to calculate
the  propagation of the source spectrum through this nebula
\citep{Groves-et-al-2008}. The H\,II region absorbs effectively all
ionizing radiation and is the source of hydrogen recombination
lines. A fraction ($f_{\rm{PDR}}$=.3) of the solid angle into which
the star particle is emitting is assumed to be obscured by the PDR,
which absorbs a significant fraction of the UV emission of the stellar
particle and reprocesses it into the FIR. The remaining solid angles
are not influenced by the PDR, and the final result is a linear
combination of the PDR of obscured and unobscured regions weighted by
$f_{\rm{PDR}}$.  Owing to the low star formation rates in the outer
disk, very few particles in the outer disk are less than 10 Myr old. 
Thus emission from stars unaffected by PDRs
dominates the FUV and the results here are very insensitive to
$f_{\rm{PDR}}$.

\begin{figure}
\plotone{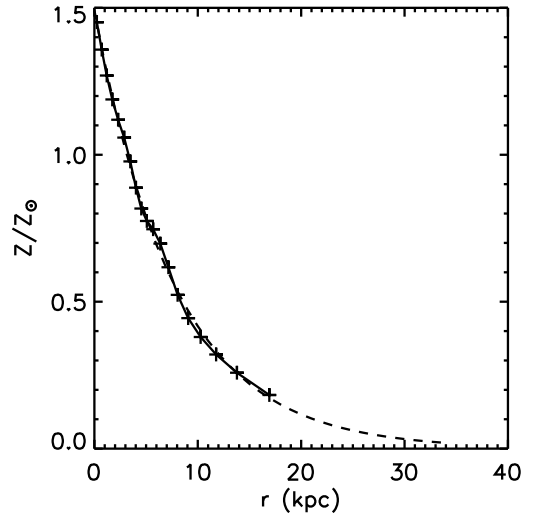}
\caption[]{Metallicity as a function of radius in our galaxy models at
  0.14 Gyr (initial snapshot). 
The solid line describes the stars and the dashed line follows the gas metallicity into the outer disk.}
\label{fig:metalic}
\end{figure}

The source spectra, as described above, are then propagated as photon
'packets' through the ISM represented by gas particles. Gas and star
particles are initialized with an exponentially decreasing metallicity
distribution with a central value of $1.6 \ \rm{Z}_{\odot}$ and a
value of $0.3 \ \rm{Z}_{\odot}$ at the break radius to approximately
match the outer disk metallicities of M\,83
\citep{Bresolin-et-al-2009}. The distribution is shown in
Figure~\ref{fig:metalic} and additional metals are added as the
simulation progresses as described in
SH03. We assume a dust to metals mass ratio
of 0.4 \citep{Dwek-1998} and use the dust models of
\citet{Weingartner-Draine-2001}, updated according to
\citet{Draine-Li-2007}. For details on the photon propagation, see
\citet{Jonsson-2006}. The final spectra are comprised of source
photons that escape absorption and scattering, supplemented by photons
re-emitted in the infrared from the dust and photons scattered into
the line of sight.  Finally, the viewing angle and bandpass are
considered to calculate broadband images. Since we examine only FUV
and \Ks-band images in this paper, dust emission is unimportant. We
only examine face-on images in this work. Highly inclined models would
have more obscuration along the line of sight and show less UV
emission.

\section{A Case Study} \label{sec:casestudy}

\subsection{Evolution} \label{sec:casestudyev}

\begin{figure*}
\plotone{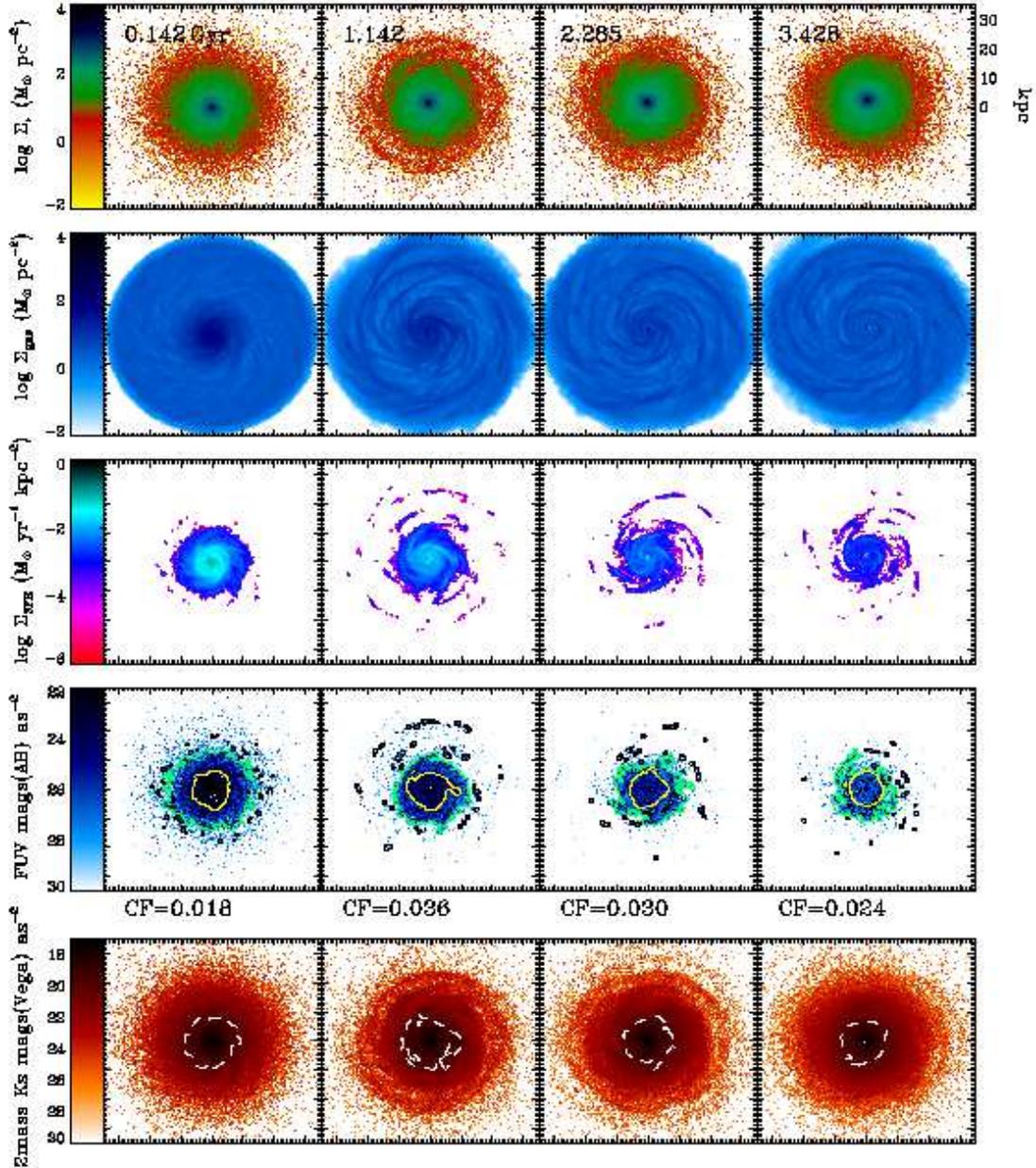}
\caption[]{Snapshots of the evolution of the flat-low
  model. Top-bottom: stellar surface density, gas surface density,
  star formation rate surface density, FUV surface brightness and \Ks
  \ surface brightness. The black contour is the 27.25 FUV AB magnitudes \as2 contour and the
yellow contour contains 80\% of the \Ks-band flux. The green contour
highlights the innermost 27.25 FUV AB magnitudes \as2 contour. The area between
the yellow and green contours is defined as the LSB zone in
T07. The CF label is the covering fraction of outer disk FUV pixels
above 27.25 FUV AB magnitudes \as2.  The white dotted
line is at 20 \Ks \ Vega magnitudes \as2.}
\label{fig:seq}
\end{figure*}

In \citet{Bush-et-al-2008} we demonstrated that spiral over-densities
in the inner disk of a galaxy can propagate to the edge of an extended
constant density gas disk. In the outer disk of that model, peaks in
the spiral waves above the density threshold form stars while 
troughs below the density threshold do not. This yields filamentary
star formation in the outer disk, as seen in Type I XUV disks (T07). A
study of the model's observable properties using radiative transfer
was beyond the scope of that work. However, the model presented there
is very similar to our flat-low model. To provide a context for the
analysis of the entire sample of simulations, we first study the
results of applying radiative transfer to the flat-low model and the
time evolution of this model's properties.

Four snapshots from the 4 Gyr evolution of flat-low are
shown in Figure~\ref{fig:seq}. The first three rows of
Figure~\ref{fig:seq} map stellar surface density, gas surface density,
and instantaneous star formation rate surface density
(top-bottom). The star formation rate, and therefore also the UV emission, is
determined by the gas distribution. Early in the simulation, ring-like
perturbations in gas density propagate outwards from the center.  These relax with
time into steady spiral features caused by potential fluctuations in
the halo \citep{Hernquist-1993}. For a discussion of how these spiral
features compare to observed spiral features in galaxies, see
\S~\ref{sec:spiralstruct}. This results in rings of star
formation in the outer disk early on, relaxing to spiral features
around 2 Gyr. Late in the simulation, the inner regions deplete as gas
is converted into stars, leading to inner holes in the gas distribution
and, eventually, also in the distribution of star formation.  The star formation
rates in the outer disk are low enough that they do not significantly
deplete the outer disk (see \S~\ref{sec:longevity}), so the morphology
of star formation in the outer disk changes very little from 2 Gyr to
the end of the simulation.

To classify disks according to their outer disk star formation, T07 analyze the FUV and 2MASS \Ks-band images of 189 galaxies in the
GALEX Atlas of Nearby Galaxies \citep{Gildepaz-et-al-2007} and define
two types of XUV disks. 
FUV emission traces star formation over $\sim$
200 Myr, not instantaneous star formation, so in order to accurately
compare our simulations to T07 we created FUV and \Ks-band images
with \textsc{Sunrise}. FUV and \Ks-band images are shown in the fourth 
and fifth rows of Figure~\ref{fig:seq}. The FUV emission
partially traces the instantaneous star formation rate, but also 
stretches 
the features out in time
as older stars continue to emit in the UV. The \Ks-band traces 
the complete stellar population, resulting in images very similar to the stellar surface density maps. 

T07 define a Type I XUV disk to be a galaxy that has at least two complexes of structured FUV emission beyond
an FUV contour of 27.25 AB magnitudes \as2 at 1 kpc resolution that are
unaccompanied by emission in the \Ks \ or DSS red bands. 
To analyze our simulations when the gas distribution is comparable to
observed spiral galaxies, we choose snapshots after the
spiral structure has become steady, 
but before the inner gas is
depleted, near the 2.285 Gyr snapshot shown. 
 At this time, knots of FUV emission trace spiral arms in the outer
disk. The features brighter than 27.25 AB mags \as2 are  highlighted
with a thin black contour in Figure~\ref{fig:seq}. The surface
brightness of the outer disk in the \Ks \ is very low. The approximate surface
brightness limit of 2MASS is 20 magnitudes \as2, shown as the white dashed
line in Figure~\ref{fig:seq} \citep{Jarrett-et-al-2003}. Most of the
emission in the outer disk is below the detection limit for
2MASS. The outer disk shows more than one knot of UV emission
without accompanying \Ks-band emission, this model most likely fits the criterion for Type I XUV
disks. We do not simulate the DSS red band, but expect the results to be similar.

T07 define Type II XUV disks based on the properties of a low surface brightness (LSB) zone in the outer disks of galaxies.  They define the LSB zone as emission beyond 80\% of the \Ks-band
flux, and interior to the innermost 27.25 FUV AB magnitudes \as2 contour. The
80\% \Ks-band flux contour for our model (80 \% of the flux
above the 20
magnitudes \as2 surface brightness limit) is shown in Figure~\ref{fig:seq} in
yellow. The innermost continuous 27.25 FUV mag \as2 contour is
highlighted in green on the same figure. The LSB zone of our
simulations is then the area between
the yellow and green contours. T07 use the color and area of the LSB
zone to define Type II XUV disks as disks whose outskirts
are very bright in the FUV, but low surface brightness in the \Ks-band. 
Their empirical definition of these objects is those
whose LSB zones have an
FUV$-$\Ks\ color of less than 4, and whose LSB zones are at least 7
times the area of the inner disk, defined by the 80\% \Ks-band
contour. 

For this model, at 2.285 Gyr, the LSB zone color is 3.6, but
the LSB zone size ratio is 2.1. Consequently, flat-low does not qualify
as a Type II disk at this time. In fact, these values are similar to
those sees in Type I XUV disks and normal disk galaxies (see \S~\ref{sec:Thilker}). 
Because star formation in
the outer disk is very filamentary, the continuous FUV contour, even at 1
kpc resolution, actually traces
the edge of the inner disk. Therefore, the LSB zone is actually
quantifying the properties of the edge of the inner disk, not the properties of the extended outer disk, so we would expect its properties to agree with those of normal galaxies.  

Since the LSB zone does not measure extended filamentary emission, we introduce a covering fraction of FUV emission to quantify the amount of outer disk emission. We define the covering fraction as the fraction
of pixels beyond the 'FUV inner disk' that are brighter than 27.25 FUV
AB magnitudes \as2 when the image is smoothed to 1 kpc resolution. 
We define the FUV inner disk to end where the average FUV surface brightness is 27.25 FUV AB
magnitudes \as2 and calculate the covering fraction between this
radius and twice this radius. The values are listed below the FUV
images. These will be discussed further in \S~\ref{sec:grid}.

\begin{figure}
\plotone{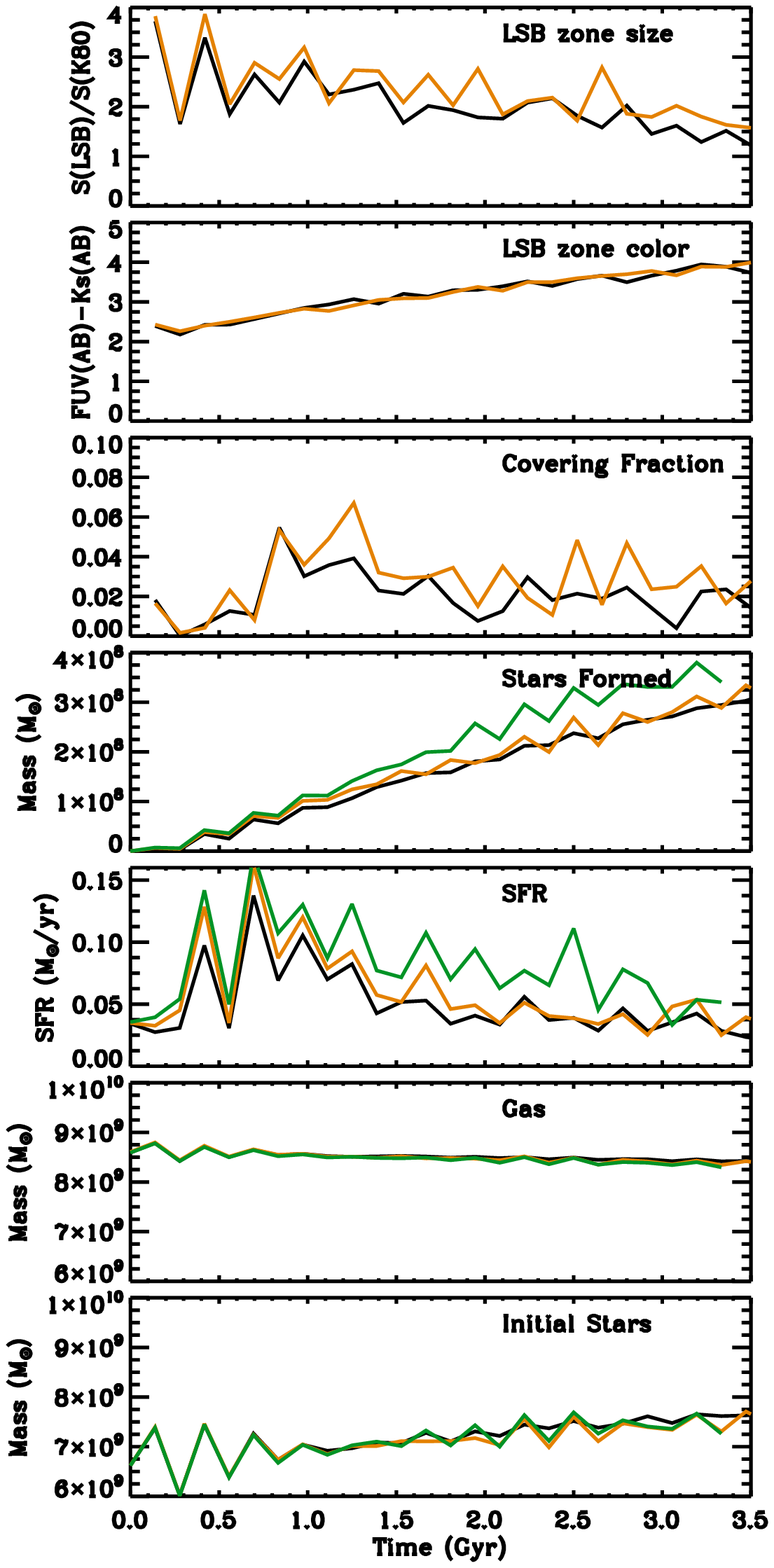}
\caption[]{Time evolution of outer disk quantities for the
  flat-low model. See the \S~\ref{sec:casestudyev} for a description. }
\label{fig:inoutevolve}
\end{figure}

To explore how the evolution of the disk affects these
classifications, we examine the properties of the outer disk with time. 
We take the
outer disk to start at a radius where the star formation rate surface
density falls below 3 $\times$ 10$^{-4}$ \Msun kpc$^{-2}$ yr$^{-1}$, the 27.25 FUV contour in star formation rate using the calibration in \citet{Kennicutt-1998-anreview}, when the simulation is initialized. The mass of stars initially in the
simulation, gas mass, star formation rate and mass of stars formed in the simulation
are summed beyond this radius every 0.14 Gyr and presented as a function of time in Figure~\ref{fig:inoutevolve}.
For the first 1.5 Gyr, as the simulation relaxes, strong oscillations are seen
in these quantities as density perturbations shift slightly inside or
outside the boundary between the inner and outer disk. After these
relax, most quantities evolve smoothly with time. The mass of old stars 
in the outer disk increases slowly over
time. This most likely owes to stellar migration, a phenomenon caused by resonant
scattering from spiral arms \citep{Roskar-et-al-2008b}. The gas mass
decreases slowly with time as it is converted into stars. The mass of stars formed increases
correspondingly. The star formation rate is high initially owing to the
strong perturbations in gas density as the disk relaxes, but then 
 slowly decreases as gas depletes. As the stellar mass of the disk builds and the star
formation rate drops, the color of the LSB zone reddens and the
size shrinks. Since Type II XUV disks have large, blue LSB zones, this means that this disk evolves \textit{away} from
being a Type II XUV disk over time. This hints that recent gas
accretion, creating in effect an un-evolved outer disk, may be important in producing Type II XUV disks. The covering fraction of FUV
emission in the outer disk is also high initially but settles at around 0.02 after 2 Gyr. It is noisier
than many of the other quantities, but evolves less
with time.

To confirm that these simulations are indeed relaxed after $\approx$ 2
Gyr, we created new initial conditions for the flat-low model by
pre-running the simulation without any star formation or gas cooling
for over 4 Gyr. We then turned star formation and cooling on and the
simulation was analyzed identically to the original. After 2
Gyr, the results are consistent with the original run. 

\subsection{Dependence on Disk Resolution} \label{sec:casestudydr}

The three observational quantities (covering fraction, LSB zone color
and LSB zone size) are
evaluated over an area that depends on the emission properties of the
disk. If the area covered by star forming particles changed
significantly with resolution, these quantities may be resolution
dependent.
 The covering fraction, with its dependence on the knots of star
formation in the outer disk, is particularly susceptible. To determine
whether these quantities depend on resolution, we ran
simulations with twice and ten times the number of particles in the gas, stars and
bulge of the galaxy. In these simulations, more gas particles of lower
mass, form stars at each time step. The results for the double
resolution run are over-plotted in
Figure~\ref{fig:inoutevolve} in orange. The observable quantities do
not change more than the scatter in the original
simulation.  Although the simulation with ten times the resolution
simulation was too large to be processed by \textsc{Sunrise}, we have
over-plotted the stellar mass, gas mass and star formation rates of
the outer disk in Figure~\ref{fig:inoutevolve} in green. Note that at
this resolution, the star formation rate increases. The simulation is now tracing smaller, but higher density
peaks in the gas density distribution. We note that this means our
results are lower limits on the star formation rate,
and most likely the covering fraction, of outer disks. However, the
qualitative results described in \S~\ref{sec:grid} are unaffected.

\subsection{Dependence on Metallicity} \label{sec:casestudymet}

\begin{figure}
\plotone{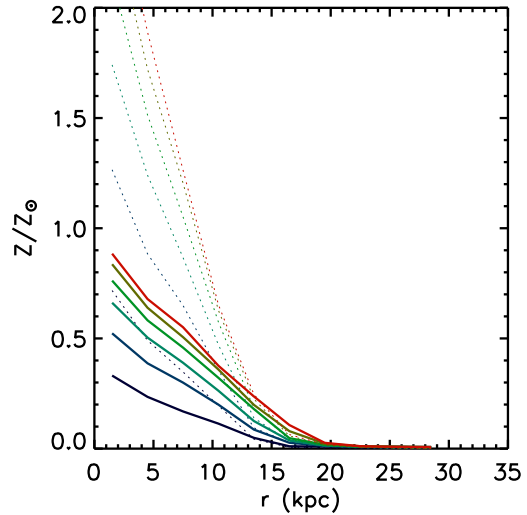}
\caption[]{Average metallicity of the flat-low model as a function of radius for the stars formed in
the simulation (solid lines) and gas (dashed lines) as a function of
time. Times are represented by colors: 0.7 Gyr - purple,
1.4 Gyr - blue, 2.1 Gyr - turquoise, 2.8 Gyr - green, 3.5 Gyr - yellow
and 4.2 Gyr - red. }
\label{fig:metalevolve}
\end{figure}

The metallicity of the stars in our simulation affects the
template SED chosen for them, and the metallicity of the gas
affects the amount of obscuration in the radiative transfer
calculations. It is therefore important to consider how the metallicity
distribution we select affects our results. The metallicity used by
\textsc{Sunrise} comes from two sources. The first is the distribution of metals present when
the simulation is initialized. Our initial distribution is shown in
Figure~\ref{fig:metalic}. The second is metal enrichment owing to
supernova feedback. As stars form in the
simulation, gas particles are treated as a closed box system, and
metals are instantaneously returned to the gas.
The new stellar particle is given the metallicity of the gas
particle from which it formed. Over time, the metallicity of the gas
and stars increases. This evolution is shown for the flat-low simulation
in Figure~\ref{fig:metalevolve}.  

\begin{figure}
\plotone{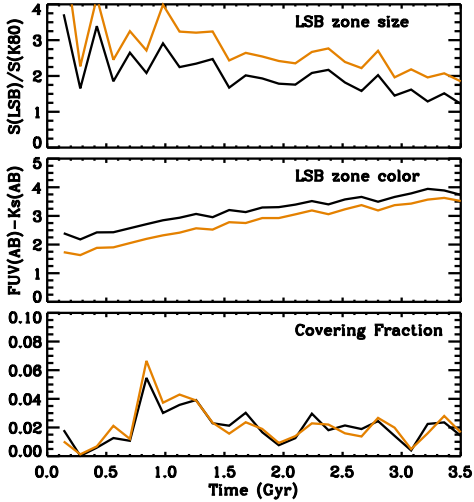}
\caption[]{Time evolution of outer disk quantities for the
  flat-low model. Over-plotted in orange are the results for a
  simulation with no initial metals. See the \S~\ref{sec:casestudyev} for a description. }
\label{fig:metaltest}
\end{figure}

Very little is known about the metallicity of outer disks. Observations
have found gas metallicities anywhere from $1-.1 \ \rm{Z}_{\odot}$
\citep{Bresolin-et-al-2009, Gildepaz-et-al-2007}. At 2.2 Gyr, the
metallicity of the new stars (the dominant source of UV emission) at
the break radius in our simulation is  $\sim 0.3 \ \rm{Z}_{\odot}$ from
supernova feedback and $\sim 0.3 \ \rm{Z}_{\odot}$ from the initial
conditions, which gives  $\sim 0.6 \ \rm{Z}_{\odot}$ overall. This is
reasonable, but could also be too high. To test how much our results
depend on metallicity, we re-ran the \textsc{Sunrise} radiative transfer on
flat-low with no initial metal content. In this case, the entire metal
contribution comes from supernova feedback. Clearly
this is a limiting case, since it also assumes that the initial simulation
stars have no metal content. The results are shown in
Figure~\ref{fig:metaltest}. The black line is the flat-low simulation
and the orange line shows the results of the radiative transfer with no initial metal
content. As we would expect, the LSB zone color is noticeably bluer throughout
the simulation, owing to the lower metallicity of the FUV emitting
stars. The LSB zone size increases owing to an increase in FUV emission
and a slight decrease in \Ks-band emission, both resulting from the
lower metallicity of the initial and new stars. Interestingly, the
covering fraction changes very little. This indicates that the change
in metallicity does not increase the brightness of knots enough for a
larger number to exceed the 27.25 FUV contour. 

While these effects are not large enough to change the
conclusions we discuss, there is some
uncertainty about the exact color and sizes of the LSB zone owing to the
lack of constraints on the metallicities of outer disks.

\section{Set of Models and Comparison to Observations} \label{sec:grid}

After the detailed examination of the flat-low model, we now study and classify the
other models in Table~\ref{tab:ics}. Spiral structure propagates from the inner to the
outer disk in all of our simulations. The amount of outer disk star
formation is then controlled by the amount of gas in the outer
disk that is above the threshold at any given time, which depends on
the density perturbations owing to the spiral structure and the radial
distribution of the gas. Immediately below, we focus on the radial gas
distribution, but we consider variations in the spiral structure in
\S~\ref{sec:spiralstruct}.

Maps of stellar surface density, gas surface density, and
instantaneous star formation rate, along with gas, stellar, and star
formation rate density profiles at 2.2 Gyr for each simulation
are provided in Appendix A. The time evolution of each model is similar
 to that seen for the flat-low model. In the following
sections, we compare our models to several observations
star formation in outer disks. 

\subsection{Comparison to Thilker et al. 2007} \label{sec:Thilker}

\begin{figure}
\plotone{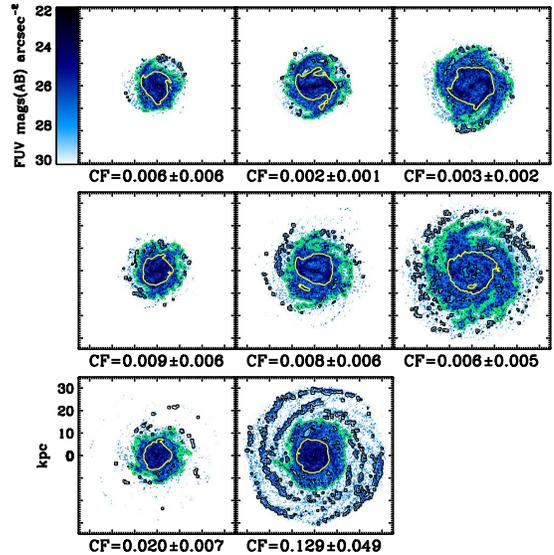}
\caption[]{2.2 Gyr FUV broadband image from each of our
simulations.  Top (L-R): exp-low, exp-mid, exp-high. Middle (L-R):
Mestel-low, Mestel-mid, Mestel-high. Bottom (L-R): flat-low,
flat-mid. Contours are described in Figure~\ref{fig:seq}. }
\label{fig:FUV}
\end{figure}

\begin{figure}
\plotone{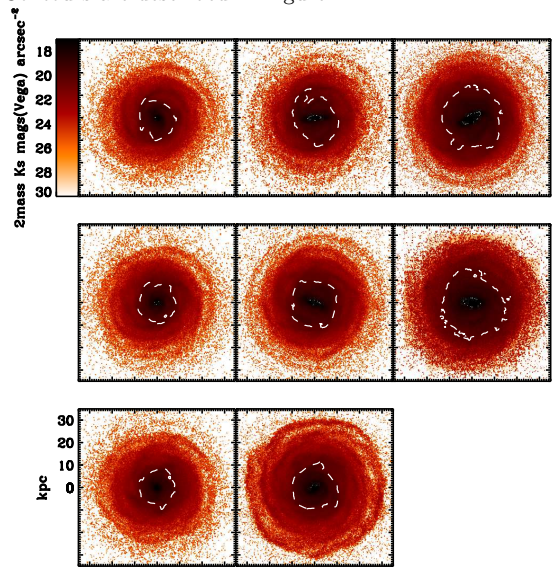}
\caption[]{\Ks-band images for the same simulations as in Figure~\ref{fig:FUV}. The white dotted
line is at 20 \Ks \ Vega magnitudes \as2.
}
\label{fig:Ks}
\end{figure}

\begin{figure}
\plotone{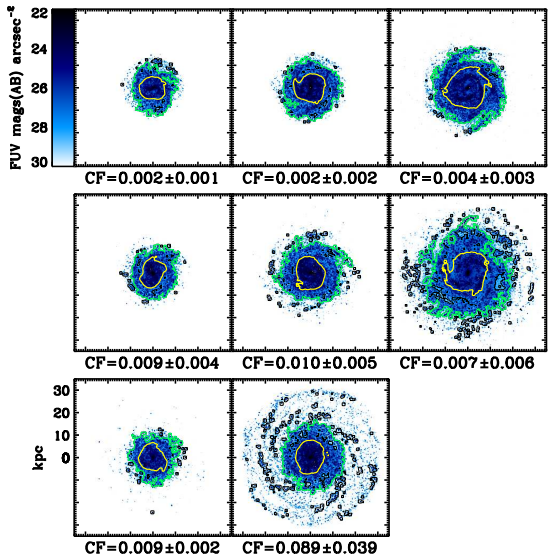}
\caption[]{ Same as Figure~\ref{fig:FUV} for simulations with
$t_{\star}$ = 6.25. }
\label{fig:FUVTsfr4.5}
\end{figure}

One snapshot from each simulation described in \S~\ref{sec:ics} is shown in
Figure~\ref{fig:FUV}. The thin black contour in these images is at
27.25 AB magnitudes \as2, and the green line highlights the innermost
continuous contour, as in Figure~\ref{fig:seq}. Clumps beyond this
green line contribute to the Type I XUV disk classification. All disks
show a few clumps beyond this contour, but how fast
the gas distribution falls off with radius controls the radial extent
of the FUV complexes. Disks with slowly varying gas distributions have
clumps of FUV emission at larger radii. The flat-low,
flat-mid, Mestel-mid, and Mestel-high models all show significant
numbers of FUV knots in
filamentary structures extending off the disk. None of the purely
exponential gas disks show more than a small 'fringe' of XUV
emission because of their rapid decrease in surface density with radius. 
The corresponding \Ks-band images are shown in Figure~\ref{fig:Ks}. The surface brightness
limits seen in Figure~\ref{fig:Ks} (white dashed contour) are at smaller radius than most of
the FUV knots seen in Figure~\ref{fig:FUV}, so the flat-low,
flat-mid, Mestel-mid, and Mestel-high models would most likely be classified as
Type I XUV disks. 

To determine how robust these results are to the star formation rate prescription
we increased the star formation timescale of the gas by a factor of
$\sim$ 3. Shown in Figure~\ref{fig:FUVTsfr4.5}, when $t_{\star}$ is increased, the overall amount of star
formation is reduced, but its distribution is largely
unchanged. However, the flat-low model and Mestel-mid model simulation may not be classified as XUV
disks, depending on the snapshot analyzed, because they have very few
knots of emission brighter than the 27.25 FUV contour.

 Based on these simulations we expect higher surface density gas disks whose outer gas profiles do
not fall off exponentially to be classified as Type I XUV
disks. Studies of H\,I distributions indicate that non-exponential gas
profiles are common in the outer parts of galaxies
\citep{Broeils-Rhee-1997, Broeils-vanWoerden-1994}. It should be noted that we only tested exponential disks with
the same scale length as the stellar disk, in this case, 3.75 kpc. As the scale length of the gas disk increases and falls off more slowly it will be more likely to be classified as an XUV
disk. 

\begin{figure}
\plotone{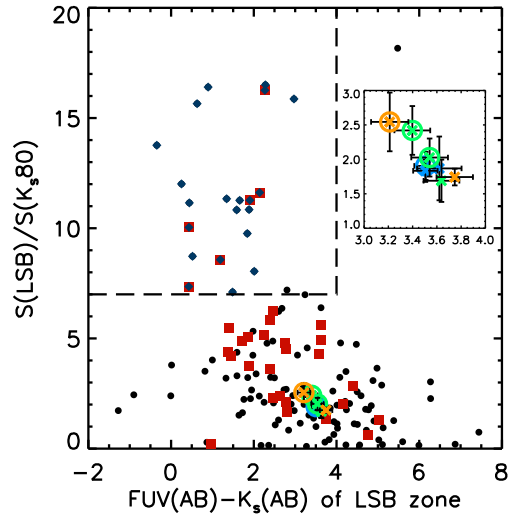}
\caption[]{LSB zone diagnostic diagram to compare to T07, with the
  ratio of the area of the LSB zone to the area containing 80\% of the
  K band flux vs. the \Ks - FUV color of the LSB zone. For a
description of the LSB zone see the text. T07 data are
  represented by black points for non-XUV disks, red squares for Type I XUV disks, and blue diamonds for Type II XUV disks. Our models are shown as crosses and models we identified as Type I XUV disks are circled. The most
high mass models are in blue, middle in green and low mass in
orange. Points are the average values over four snapshots and error
bars are the standard deviation of these values.   Our models are consistent
with being normal or Type I XUV disks. The inset shows the detail of the region where our models lie.}
\label{fig:type2eval}
\end{figure}
\begin{figure*}
\plottwo{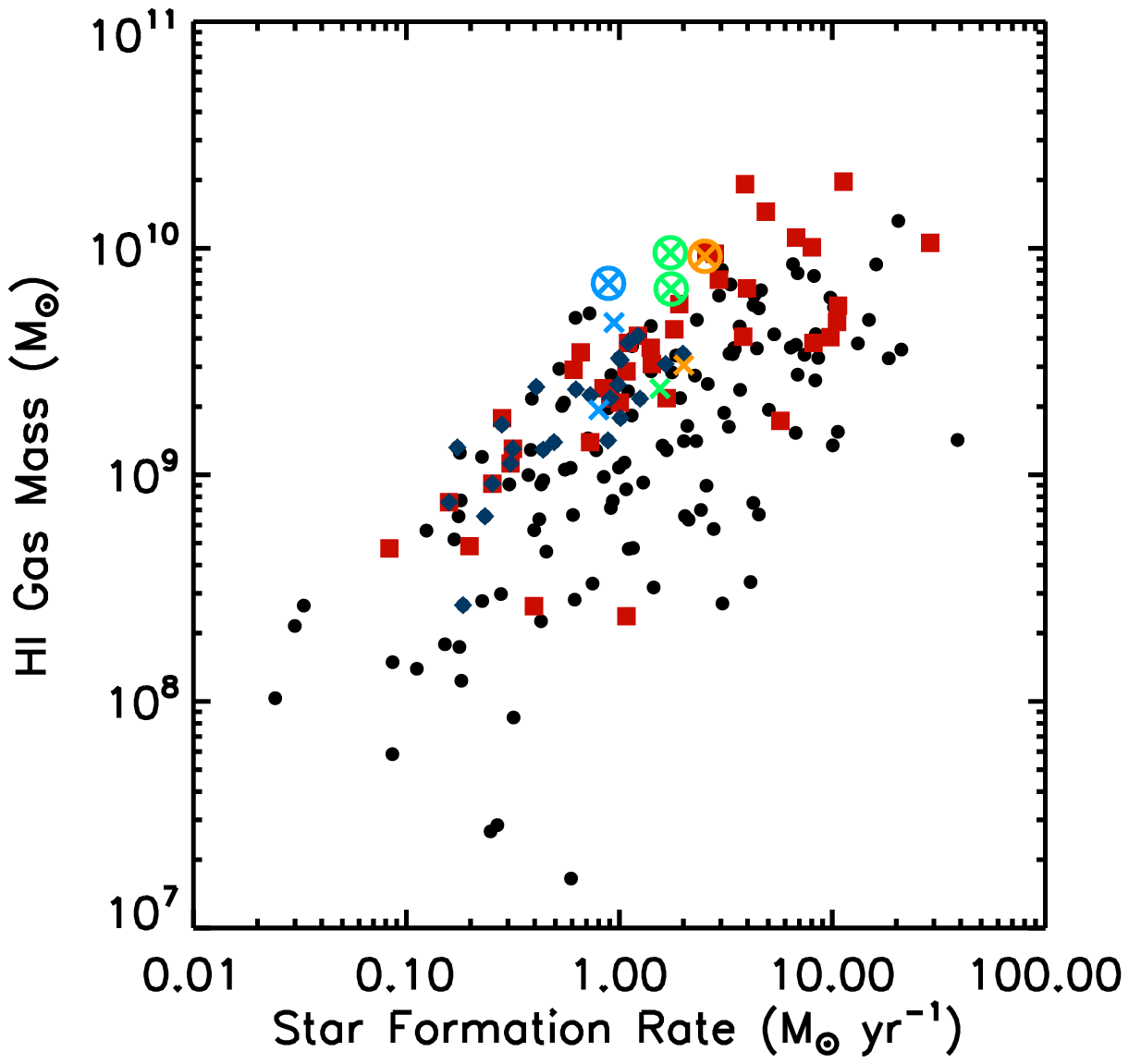}{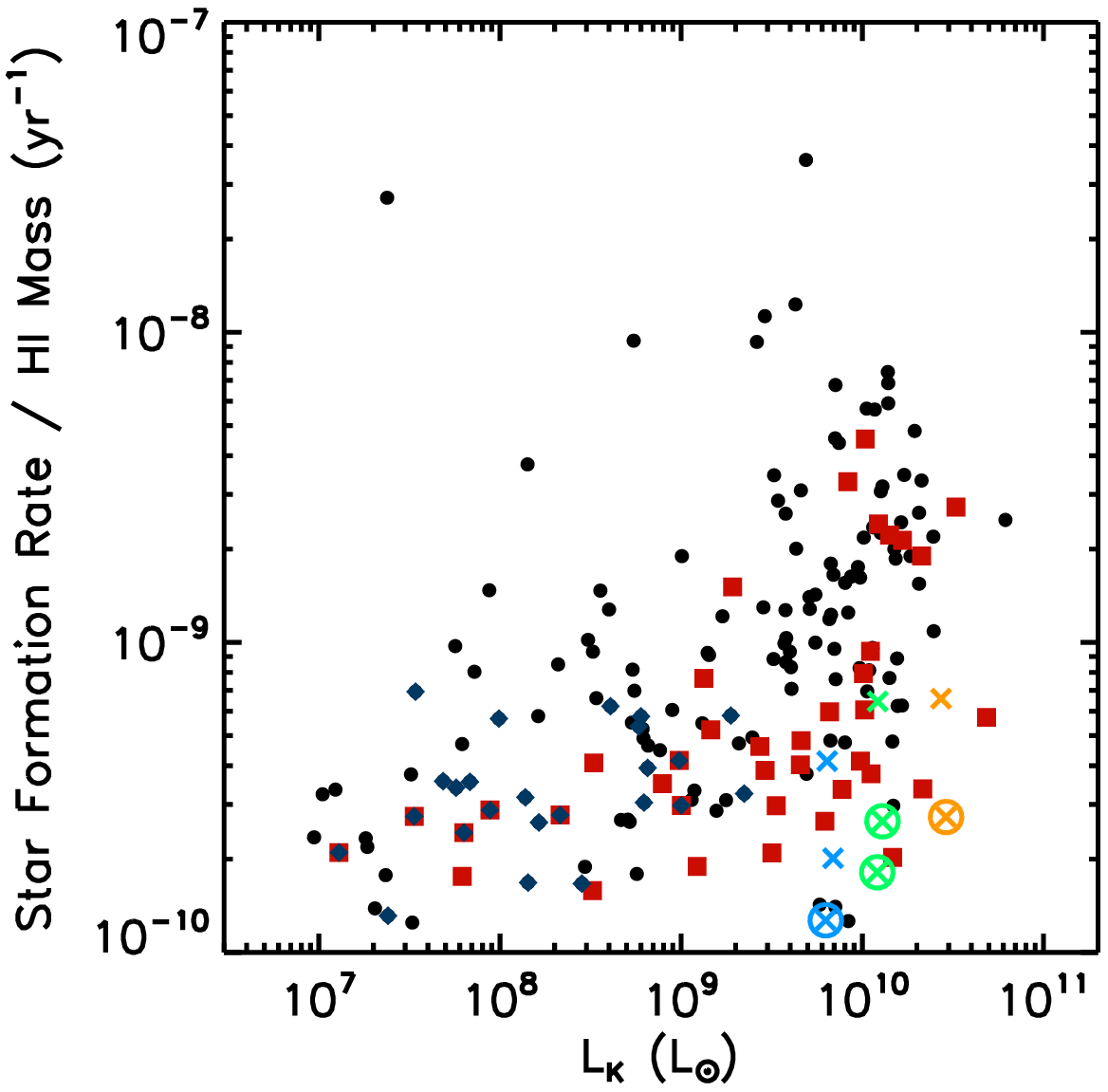}
\caption[]{Left: SFR vs. H\,I mass for our simulations at 2.2 Gyr
(symbols as in Figure~\ref{fig:type2eval}) compared to T07. Our models
are consistent with normal and Type I XUV disks. Right: SFR/H~I
vs. $L_{k}$ for our simulations at 2.2 Gyr (symbols as in
Figure~\ref{fig:type2eval}). Our models classified as Type I XUV
disks appear to lie in a region mostly populated by Type I XUV disks.}
\label{fig:thilkerfigs}
\end{figure*}

To determine 
whether any of the models would be classified as a Type II disk, we evaluate the
color and area of the low surface brightness (LSB) zone defined by
T07. As described in \S~\ref{sec:casestudyev}, T07 define the LSB zone as emission beyond 80\% of the \Ks-band
flux, and interior to the innermost 27.25 FUV AB magnitudes \as2 contour. The
80\% \Ks-band flux contour for the grid of models is shown in yellow
in Figure~\ref{fig:FUV}. The LSB zone of our grid of simulations is the area between
the yellow and green contours in Figure~\ref{fig:FUV}. T07 define Type II XUV disks as objects whose LSB zone has an
FUV$-$\Ks\ color of less than 4, and the LSB zone is at least 7
times the area of the inner disk, defined by the 80\% \Ks-band
contour. The color vs. area of this zone for our sample are shown in
Figure~\ref{fig:type2eval} with the T07 sample.  

None of our simulations meet the Type II criterion. Although they are
blue enough in FUV-\Ks, the LSB zone is too small. These
characteristics are consistent with normal and Type I disks.  Note
that our simulations have very small scatter in this plot. As with the
discussion of the flat-low model, the outer
disk star formation in our simulations covers such a low fraction of
the disk that the innermost continuous 27.25 FUV AB magnitudes \as2 contour
tends to fall at the edge of the inner disk, not the outer extended
gas disk, so the LSB zone is primarily probing a part of the disk where the initial
structure of each model is very similar. By varying the structure of the inner disk, particularly
increasing the scale lengths of the models, we could change the
characteristics of the LSB zone. However, it seems unlikely that
we could alter the structure of the inner disk enough that it would yield Type
II disks, while staying consistent with the observed structure of
galaxies. 

To quantify the outer disk UV emission we calculate covering fractions as
described in \S~\ref{sec:casestudyev}. Since the covering fraction
varies significantly between snapshots, we considered 4 snapshots spanning an evolutionary time of
2.1 Gyr to 2.5 Gyr and report their average value in
Figure~\ref{fig:FUV}. The error reported is the standard deviation of
these values. Overall, the covering fraction seems to
quantify the amount of FUV emission well, but, especially when
limited to such a small sample, the covering fraction does not
differentiate strongly between galaxies with different gas
distributions. On average, the Mestel models have a higher covering
fraction than the exponential models, and likewise for the constant
density models vs. the Mestel models. But the covering fraction is not
significantly higher for the models classified as Type I XUV disks than for
the others. This at least partially owes to 
resolution. The contours are smoothed to 1 kpc in all
models, the area of the low mass disks is smaller, increasing the size of the UV knots in the outer disk relative to the area the covering fraction is calculated over. This biases the
covering fractions to be larger for the lower mass disks.  

To further constrain our models, we compare their properties to other
observational characteristics of XUV disks. T07 show that XUV disks
are gas rich for their total star formation rate. In
Figure~\ref{fig:thilkerfigs} we plot the approximate H~I gas masses of
our disks against their total instantaneous star formation rates,
along with the T07 sample. The H~I gas mass for our simulations was
approximated by assuming that 50\% of all the star forming gas is
molecular \citep{Keres-et-al-2003} and that 70\% of what remains is H\,I (to
account for He).  Our models all lie in the high gas mass region of the
T07 data, unsurprising since we have chosen a massive galaxy
as our fiducial model. Additionally, in star forming regions, the molecular gas percentage
may in fact be much higher, lowering the H\,I masses.
However, since they are high mass disks, all of our galaxies lie in
areas where there are both Type I XUV disks and normal disks. So while
our results are consistent with T07, this plot does not provide a good
diagnostic for determining whether our models match observed XUV disks
in terms of gas richness.

In Figure~\ref{fig:thilkerfigs} we also plot the SFR/H\,I ratio
vs. the \Ks-band luminosity (calculated inside the 20 magnitudes \as2
contour) of our galaxies, and compare these to T07. This plots gas richness vs.  L$_{\rm{K}}$, a proxy for
stellar mass and show whether the disks are gas rich for their
star formation rate and stellar mass. Our modeled Type I XUV disks fall in an area that is dominated by Type I XUV disks, rather than normal galaxies. This is promising, indicating that the
structure of our model galaxies is similar to Type I XUV disks.

In each of these plots it is evident that our models with constant density
extended gas disks tend to have slightly larger H\,I masses than most
observed galaxies. This is unsurprising; in reality we would expect to
find few galaxies whose H~I disks do not fall off at all with
radius. However, they are an interesting limiting case. Our models with
$r^{-1}$ extended disks seem to be quite consistent with the H\,I
masses of massive XUV disks. Overall, the structure of our models that are classified as Type I XUV disks are consistent with XUV disks observed by T07.
 
It is difficult to determine
the fraction of XUV disks that owe to isolated in situ star formation
if we do not know the fraction of XUV disks that have disk structures
similar to that of our Type
I XUV disk models. Detailed studies of the H\,I profiles of XUV disks
are needed to determine whether 
we are describing them fully or if there are additional conditions
needed to create all Type I XUV disks.

\subsection{Comparison to Boissier et al. 2007} \label{sec:Boissier}

\begin{figure*}
\plottwo{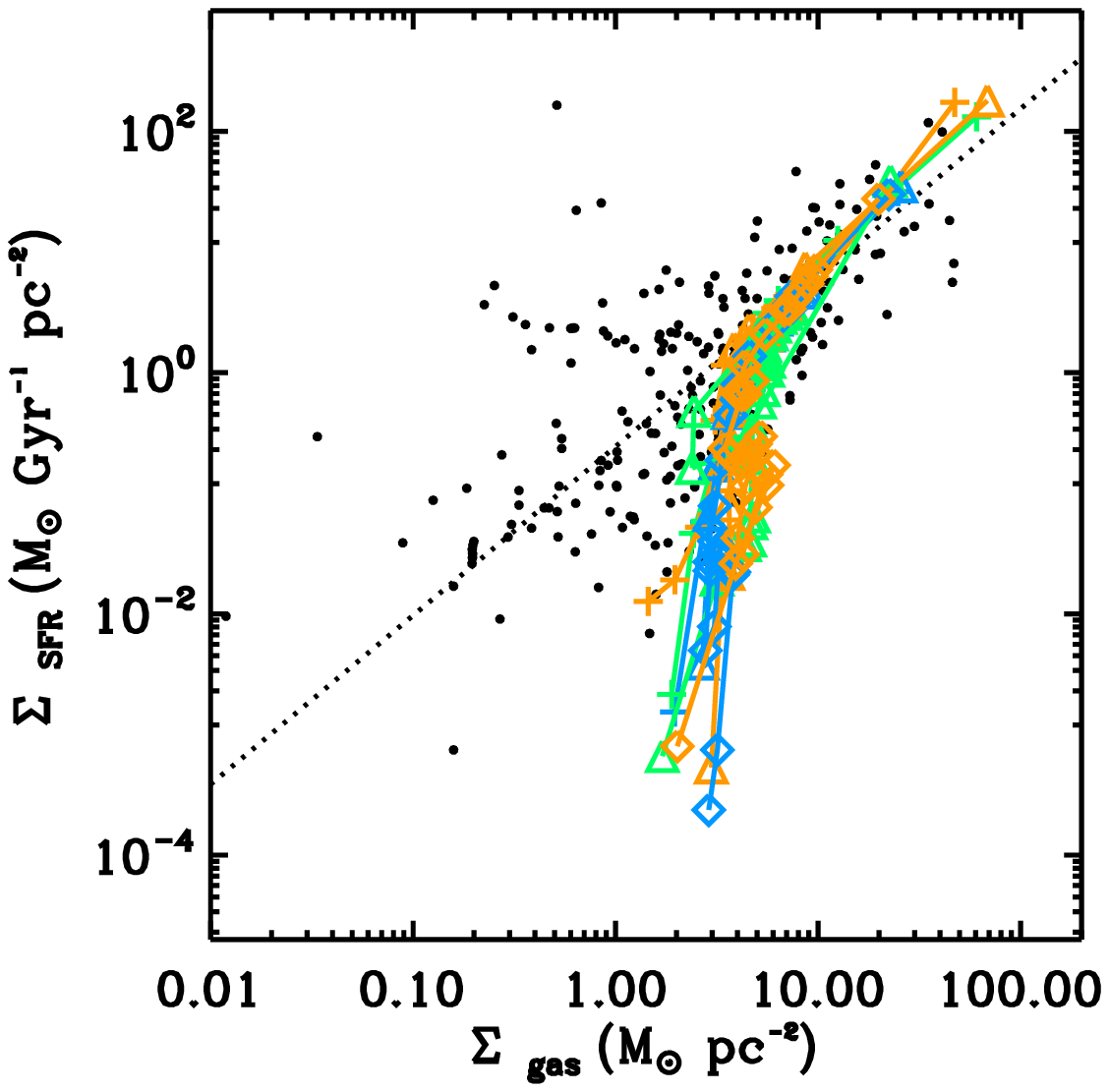}{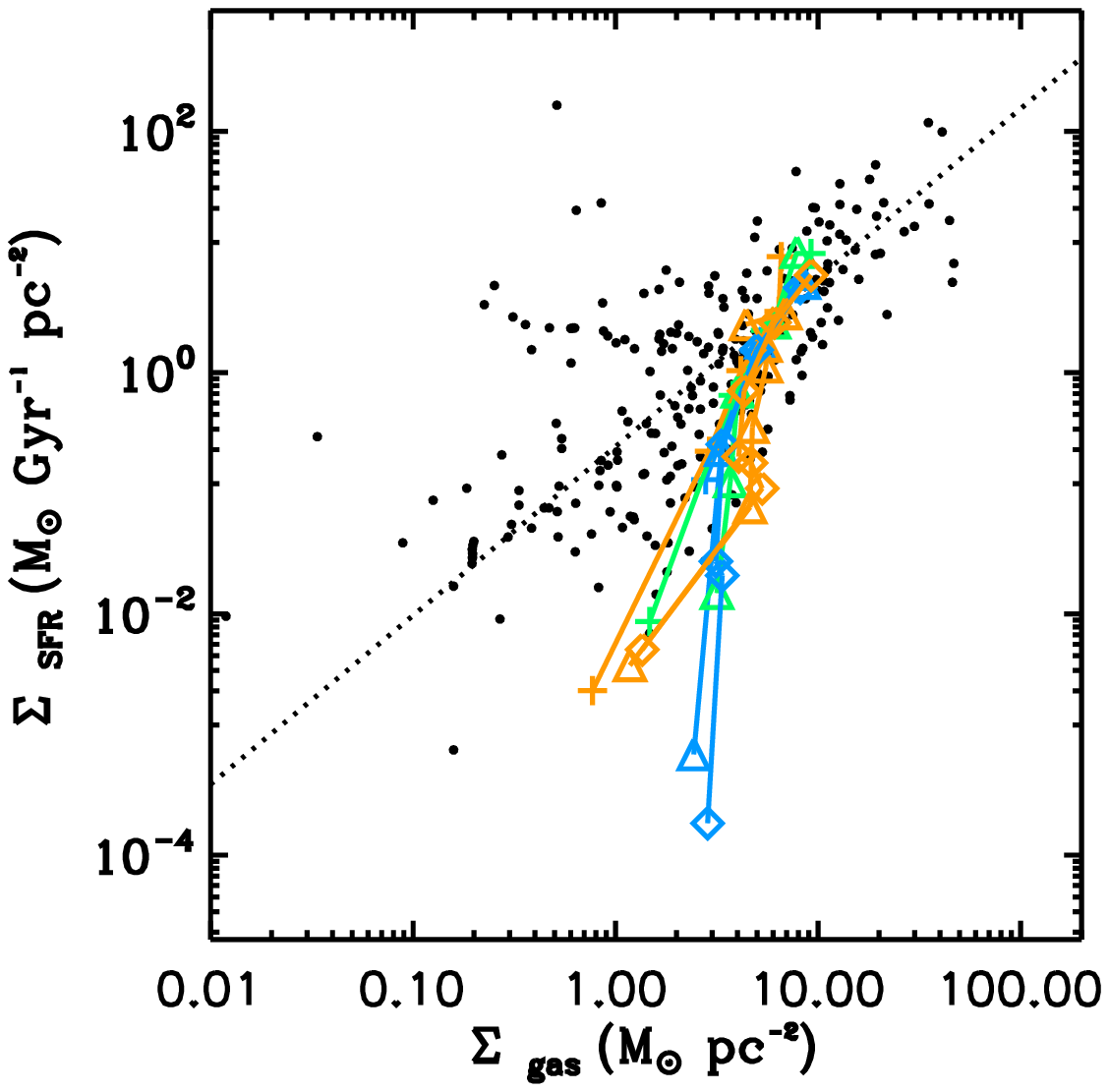}
\caption[]{Plots of star formation rate vs. gas surface density
radially  binned for all our models:
high surface density in blue, middle surface density in green and
low surface density in orange. Outer disk profiles are separated by
symbols: exponential - plus signs, Mestel - triangles and constant
density - diamonds. On the left, quantities are averaged in radial bins
1 kpc wide and on the right 5 kpc wide. While this can change the
detailed nature of the profile, it does not eliminate the truncation
in the profile.  The dashed line is the empirical law defined by
\cite{Kennicutt-1998}.  Data from B07 are plotted as black points.}
\label{fig:kennSH03}
\end{figure*}

B07 studied the extinction corrected UV radial profiles of 43 galaxies
which were selected for large angular size. In contrast to
studies of \Ha\ profiles of galaxies, which find a strong truncation at
3-5 \sdunits\ in total gas surface density \citep{Kennicutt-1989,
Martin-Kennicutt-2001}, B07 find that star formation rate profiles
derived from UV profiles do not show a break.  In
Figure~\ref{fig:kennSH03} we show the instantaneous star
formation rate of our simulations vs. the gas surface density. UV
derived star formation rates give the star formation rate averaged
over 200 Myrs, but since the star formation rates of our simulations
are mostly constant with time after relaxing initially, the trends
seen should be identical. 

While
at low gas densities, simulations can differ considerably in the
details, all simulations, including Type I XUV disks, show strong
truncations as a result of the volume density threshold to star
formation and azimuthal averaging. The levels of star formation at low gas
densities indicated by B07 are simply not reflected in our
simulations. However, there are large uncertainties in the low star
formation rate measurements and, when a gas profile is slowly varying,
averaging can play a significant role in smoothing the star formation
profile. To demonstrate this we plot the gas density vs. star
formation rate density radial profile with radial bins of two different sizes, 1 kpc and
5 kpc. While with larger bins the profiles do flatten, they still
show significant truncation. If the B07 results hold true with more
detailed studies of individual objects, a change in the star formation
prescription should be needed at low gas surface densities for at
least some galaxies. 

If the resolution of our simulation was
increased, it is possible it could trace smaller overdensities and
increase the amount of star formation at low average gas
densities. However, when the azimuthally averaged Kennicutt-Schmidt
law is plotted for the flat-low simulation at ten times the disk
resolution, a truncation is still very evident. 

\subsection{Comparison to Dong et al. 2008}

\begin{figure}
\plotone{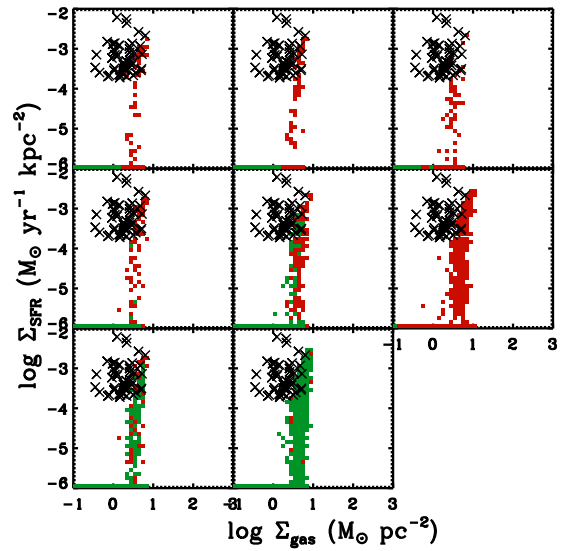}
\caption[]{Plots of star formation rate vs. gas surface density in
$1 \times 1$ kpc apertures over the outer disk. Green points indicate star formation at that rate and surface density 
beyond approximately 1.5 optical radii as well as between the optical radii and 1.5 optical radii. Red points indicate star formation  and
are interior to 1.5 optical radii. Models are the same as in
Figure~\ref{fig:FUV}.  The over-plotted black crosses are measured SFRs and H\,I gas
densities (lower limits on the total gas density) from the outskirts
of M\,83 \citep{Dong-et-al-2008}.}
\label{fig:localkenn}
\end{figure}

\citet{Dong-et-al-2008} studied the characteristics of UV selected
sources in the outer disks of a prototypical Type I XUV disk, M\,83, using
circular apertures with a diameter of 440 - 600 pc.  They found that
most of the star forming complexes are coincident with H\,I gas 
above the Toomre dynamical instability threshold. In addition, using
UV and 8 \mum derived star formation rates and plotting these against
H\,I gas density, they found general agreement with the
Kennicutt-Schmidt law derived for regions in the inner disk of M\,51
\citep{Kennicutt-et-al-2007}. 

In order to compare to this type of analysis, we divide the star
formation rate and gas surface density maps of each simulation into
1 kpc $\times$ 1 kpc squares and plot the star formation rate
vs. gas surface density for squares in the outer disk in
Figure~\ref{fig:localkenn},  along with the \citet{Dong-et-al-2008}
data. The SPH smoothing length is typically 1 kpc for our outer disk
particles, therefore these are the smallest apertures we can use and they
should be compared to the smaller aperture data of
\citet{Dong-et-al-2008} with caution. With the exception of a few points (and \citet{Dong-et-al-2008}
caution that the uncertainties on the 8 \mum portion of the star
formation rates are large), we create star formation rates as high in
the outer disk as those measured in M\,83. Some gas surface densities
in the \citet{Dong-et-al-2008} data are considerably lower than our
simulations, but they only include
H\,I, so these should be considered lower limits on the gas surface
density. If we calculate the H\,I surface density for our simulations
assuming 50\% of all star forming gas is molecular (clearly not
accurate on large scales in outer disks but in the small apertures
used by \citet{Dong-et-al-2008} this may be a good approximation) we
find that the central concentration of the M\,83 data agrees well with
our simulation results. In addition, some of our star formation rates
are lower than those see in in \citet{Dong-et-al-2008}, but this is
expected because they lay apertures only on sources, while we simply
bin our outer disk, so some of our bins will have very low star
formation rates. 

As the star formation prescription is the same in each model, the
Kennicutt-Schmidt law averaged on local scales change very little as the
disk structure is changed between models. 
The data agree similarly with 
all the models we considered. The main difference
between the models with and without Type I XUV emission is the
extent of the star formation. To highlight this, we have colored
portions of the diagram where star formation exists beyond a distance of approximately 1.5 optical radii 
in green. It
is clear that the exponential disks do not show any star formation at
large radii and, at large radii, only Mestel-mid, flat-low and flat-mid agree with the data.

\section{Discussion} \label{sec:discuss}

\subsection{Spiral Structure} \label{sec:spiralstruct}

To determine whether we are accurately reproducing the star formation
observed in galaxies, we need to determine whether the spiral
perturbations in our simulations are consistent with those in actual
galaxies. Though we know that spiral structure in our simulations is
caused by numerical noise in the halo \citep{Hernquist-1993}, what
causes spiral structure in nature is unknown. However, for our
purposes we only need to
accurately recreate the amount of gas above the star formation
threshold and therefore the amplitude of the spiral perturbations in
the outer disk.

\begin{figure}
\plotone{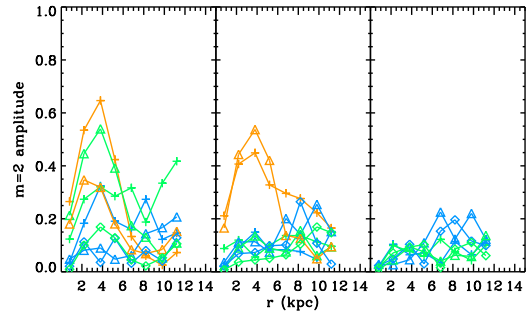}
\caption[]{The m=2 spiral mode amplitude for our models where the halo to disk mass
 ratio increases left to right. Disk
 mass fractions are 0.041 (left), 0.02 (middle) and 0.01 (right) of M$_{\rm{tot}}$. Colors and symbols
 correspond to those in Figure~\ref{fig:kennSH03}.}
\label{fig:spamp}
\end{figure}

Studies of the spiral structure in stellar disks indicate that the two
armed mode of spiral structure ($m=2$) dominates in most galaxies. We
plot the relative amplitude of the stellar $m=2$
mode for the 2.2 Gyr snapshot of each simulation on the left in 
Figure~\ref{fig:spamp}. The very high amplitudes in inner regions owe to
central bars. These are generally consistent with the range
of observed $m=2$ mode amplitudes
\citep[][Kendall et al. in preparation]{Rix-Zaritsky-1995,
Elmegreen-et-al-1992}. The stellar mass dominates in the inner disk and
gas falls into the potential well of the stellar spiral arms. These arms then propagate in
the outer disk as sound waves. Since we reproduce the amplitudes of the stellar spiral
wave correctly, it is reasonable to believe that we are accurately
representing the gas perturbations as well.

To explore how the star formation in outer disks is affected
by variations in spiral structure, we re-run our models again after 
increasing the halo mass by factors of two and four. As the halo to disk mass ratio
increases, the disk becomes less self-gravitating and the power in the
$m=2$ mode is decreased, making the spiral structure more
flocculent. The structure of the disk is kept constant, and the number
of particles in the halo is increased  to keep the
mass per particle in the halo constant between models. For the most massive disks,
increasing the halo by four times means that the rotation curve peaks at
over 300 km s$^{-1}$, which is unreasonably high, so these models are not
included in the analysis. The $m=2$ mode amplitude for the simulations
with higher halo to disk mass ratios are shown on the middle and right panels
in Figure~\ref{fig:spamp}. Note that as the halo to disk mass ratio increases, the
$m=2$ mode amplitude decreases. These are still well within the
constraints of observations \citep[][Kendall et al. in preparation]{Rix-Zaritsky-1995,
Elmegreen-et-al-1992}. 

\begin{figure}
\plotone{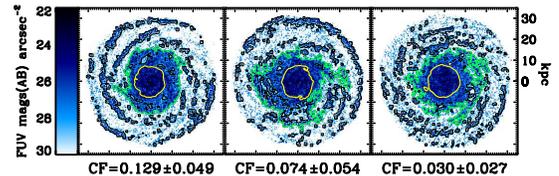}
\caption[]{FUV images of the flat-mid run as the halo
  to disk mass ratio is increased by a factor of two (middle) and four
  (right). Contours are as in Figure~\ref{fig:FUV}.
} 
\label{fig:FUVFD}
\end{figure}

\begin{figure}
\plotone{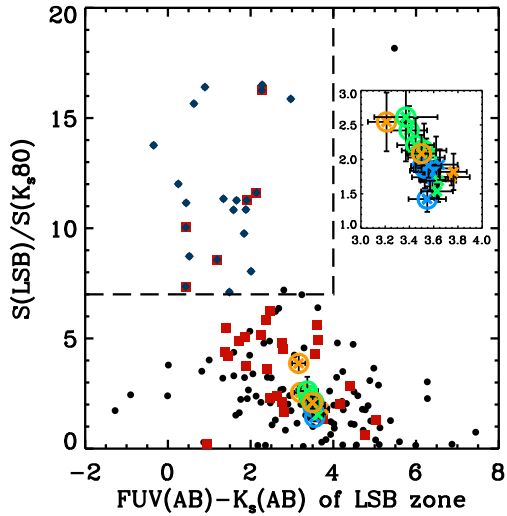}
\caption[]{The properties of the T07 LSB zone for all models, including high halo mass models (see
\S~\ref{sec:spiralstruct}). Note that the locus of points is similar to
Figure~\ref{fig:type2eval}. Colors and symbols are the same as Figure~\ref{fig:type2eval}.
} 
\label{fig:type2evalFD}
\end{figure}

An example of the effect on the morphology of the outer disk star
formation is shown in Figure~\ref{fig:FUVFD}.  Figure~\ref{fig:FUVFD}
shows the 2.2 Gyr snapshot for the flat-mid model as the halo to disk
mass ratio is increased.
The morphology is affected and this is reflected in the covering
fraction. This indicates that in galaxies
that are more dark matter dominated, such as dwarf galaxies, we might
expect to see more flocculent spiral structures in the outer disk and
smaller covering fractions. 
There is very little difference in other diagnostics.
Plotting the star formation rate vs. gas surface density does not
alter the results seen in Figure~\ref{fig:kennSH03}. In Figure~\ref{fig:type2evalFD} we show the T07 XUV type diagnostic
diagram including the high halo mass models. The locus
of points is almost entirely unaffected by the additional models.

Changing the mass resolution of the halo particles also modifies the spiral
structure. Higher resolution in the halo leads to less numerical
noise and suppresses spiral structure. We choose our halo
resolution because it recreates the observed
amount of spiral structure. 

\subsection{Stellar Profiles}

In a study of a large sample of SDSS galaxies,
\citet{Pohlen-Trujillo-2006} found that only 10\% of galaxies have a
purely exponential stellar profile.  Most galaxies show
either a ``down-bending break'', where the exponential profile of the
outer disk is steeper than the inner disk, or an ``up-bending break'' where the exponential profile of the outer
disk is shallower than that of the inner disk. While several other
methods of populating the outer disks of galaxies have been proposed
\citep{Roskar-et-al-2008a, Roskar-et-al-2008b, Younger-et-al-2007}, older
stellar populations accompanying the low levels of star formation in outer disks have been
suggested as an explanation for up-bending breaks \citep{Thilker-et-al-2007, Elmegreen-Hunter-2006}. The stellar
profiles of our models, shown in Appendix A, exhibit features owing to over-densities, but generally appear to be
characterized by a single exponential. 
The overall level of star formation in outer disks is low, as
shown by the star formation rate profiles in Appendix A, so it is
not surprising that the outer stellar profile changes very
little. 

\citet{Elmegreen-Hunter-2006} present a model of star formation
which creates down-bending breaks.
These build stellar populations which also show a down-bending
break. Their model includes a threshold for star formation that depends on the
Toomre stability criterion and a turbulent model of the ISM. Because
our threshold for star formation is just a single volume density, the
star formation profile of our simulations is determined entirely by
the gas density profile of the galaxy. Our star formation rate profiles
then flatten in the outer disk following the flattening extended
disk, rather than falling off with a sharper exponential. A more complex
implementation of the star formation threshold is needed in
order to determine whether our models agree.

\subsection{Ionized Gas}

We construct the gas content of our model galaxies to reproduce observed H\,I
distributions of galaxies.  However, there may be a significant amount
of diffuse ionized gas, particularly in the outer regions of galaxies, that is not captured in 21 cm observations that
measure the H\,I content. In
the Milky Way the
column density ranges from 20\% -60\% of the
column density of 
H\,I \citep{Reynolds-1991}.  
However, in outer disks,
low gas densities reduce the efficiency with which the gas is able
to shield itself from the UV background
\cite[e.g.][]{Elmegreen-Parravano-1994,Schaye-2004}.
Studies of H\,I line-widths with radius indicate that there is an
unexplained source of heating in outer disks
\citep{Tamburro-et-al-2009} which could also point to a contribution
from the UV background.  This indicates that H\,II could become a
significant component of the ISM at large radii. If star formation
truly does reflect the total gas mass, as it would if it
depended on the 
mid-plane pressure of the ISM, the pressure owing to the H\,II mass could boost star
formation rates in the outer disk. In the case of our simulations,
adding H\,II would increase the amount of gas above the
threshold in the outer disk and result in more star formation. Detailed studies of the ISM in
outer disks are needed to determine whether this effect is
important.

\subsection{Longevity} \label{sec:longevity}

\begin{figure}
\plotone{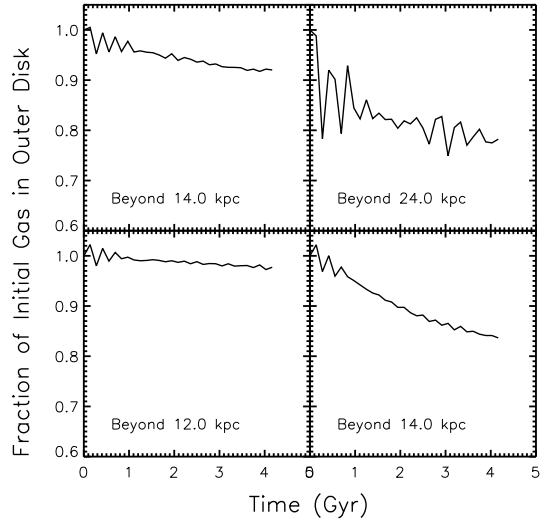}
\caption[]{Fraction of initial gas still present in outer disk
 as a function of time for the four simulations showing Type I XUV
emission. Top Row: Mestel-mid and Mestel-high. Bottom: Flat-low and flat-mid.}
\label{fig:gascontent}
\end{figure}

There is much debate over whether star formation in outer
disks is continuous or episodic. In many examples there is enough H\,I
gas to support continuous star formation at the levels seen over the
lifetime of the galaxy \citep{Gildepaz-et-al-2007}. Our simulations
are consistent with this. In Figure~\ref{fig:gascontent} we show the
fraction of the initial gas in the outer disk still present as a
function of time for the four simulations that are identified as Type I XUV
disks in \S~\ref{sec:Thilker}. The outer disk
is defined to start where the star formation rate surface
density falls below 3 $\times$ 10$^{-4}$ \Msun kpc$^{-2}$ yr$^{-1}$ in the initial disk. The oscillations
owe to non-steady behavior in the disk as the
simulation relaxes (see Figure ~\ref{fig:inoutevolve}). Even the simulations with
the most outer disk star formation, flat-mid and Mestel-high,
lose only $\sim 20$\% of their gas over the 4 Gyr timeframe of the
simulations. Flat-low and Mestel-mid show almost no change over this
time. This indicates that XUV disk emission could continue for the
lifetimes of the galaxies, though as the gas is consumed in the
galaxies with higher levels of outer disk star formation, the amount of XUV
emission may decline to what is seen in lower gas content models.

\subsection{Interactions}

In this paper, we study only isolated galaxies, where density
perturbations in the outer disk result from spiral structure
induced by numerical noise in the halo.  Of course, other sources, such
as interactions, can cause perturbations in outer disks leading to in situ star formation. Strongly shocked gas owing to interactions
in the outer disk would likely create high levels of star formation
in very small regions.  We will explore this
possibility in due course.

\subsection{\Ha \, Emission}

While exploring the simulated H$\alpha$ colors and distributions 
of individual star clusters would be very interesting, our resolution limits us to
particle masses of $\sim10^{5}$ \Msun. Since \Ha \ knots in the outer disk are 
often illuminated by a single star emitting ionizing radiation, we do not have 
the resolution to study the \Ha \ properties of our sources. 

\section{Conclusions} \label{sec:conc} 

We have explored how outer disk star formation in extended gas disks
resulting from a
Kennicutt-Schmidt star formation prescription and volume density
threshold compares to observations of XUV disks. We used a set of models with outer disk
structures motivated by H\,I surveys. We find that models with slowly varying
high surface density outer gas disks would be
classified as Type I XUV disks. 
Possible variations in the star
formation rate normalization or spiral structure amplitude
 mean that gas distributions that result in only a few
knots of outer disk UV emission in our models may not always have UV
emitting sources in their observed outer disks. While we cannot rule out
other explanations 
for Type I XUV disks, since the Type I XUV
emission in our models results naturally from empirical star formation
laws and the observed structure of galaxy disks, we argue that the theory outlined here is a viable explanation. 

The star formation in the
outer disk of our models is not prevalent enough to erase the tell-tale truncation
of the profile caused by the star formation volume
density threshold. Since B07 see no truncation in the UV with gas
radial profile, this could indicate that we need to revise our star
formation prescription. 
However, the star formation rates in our models are consistent with
star formation rates in the Type I XUV disk M\,83 \citet{Dong-et-al-2008}. More detailed
observational studies of the relation of gas density to star formation
rate in outer disks are needed to
determine whether the observed fields of M\,83  are a special case or
whether disks selected for XUV emission
are subject to different processes than the B07 sample. 

While XUV emission does not depend enough on the detail
of gas density profiles to distinguish easily between them,
it is a good indicator of the presence of
gas at large radii in galaxies. Our results imply that nearly 1/3
of galaxies have large amounts of gas at 2-3 times their optical
extents. 
In addition, forming Type II XUV disks  with traditional star
formation laws must require some process, such as rapid gas accretion,
for creating a large young stellar population without an accompanying older
population. These results may be
outlining a picture where accreting gas in the
outer parts of disks is common at late times, as suggested by the
inside out disk formation scenario demonstrated by simulations of
structure formation 
\citep[e.g.][]{Mo-Mao-White-1998, Brook-et-al-2006}, and
supported by the possibility that cold clouds developing in
cosmological simulations may accrete onto disks at late
times \citep{Keres-Hernquist-2009}.

The next step in exploring the
observed signatures of in situ star formation would be to run very high resolution
simulations that could be compared to observational work such as that of
\citet{Herbert-Fort-et-al-2009}. These authors have computed the cross
correlation of young clusters in NGC 3184 and found a significant
inter-cluster signal for separations of less than 1 kpc, with cluster
masses of $\sim 10^{3}-10^{4}$ \Msun. With higher resolution simulations, it
is possible that measurements like these could differentiate between
in situ star formation and stellar migration scenarios.  Moreover,
simulations with star formation prescriptions that are linked directly
to the molecular gas may help reconcile in situ star formation and the
results of B07.  \citet{Robertson-et-al-2008} present models
indicating that directly relating the star formation rate to the molecular
gas may eliminate truncations in the star formation rate
profile. Continued observational studies and sophisticated simulations are needed
to explore this complex regime in star formation.

\acknowledgements We thank Robert Kennicutt, Dusan Keres, Zhong Wang, Brant Robertson
and Jacqueline van Gorkom for useful conversations in the preparation
of this work.  Special thanks go to Sarah Kendall for help in the spiral structure analysis and Samuel Boissier for allowing us to use his data. 
Support for TJC was provided by the W.M. Keck
Foundation. CCH acknowledges support from a National
 Science Foundation Graduate Research Fellowship.
The computations in this paper were run on the Odyssey cluster supported by the FAS Research Computing Group.
\bibliography{xuv_v5}

\appendix

\begin{figure}
\plotone{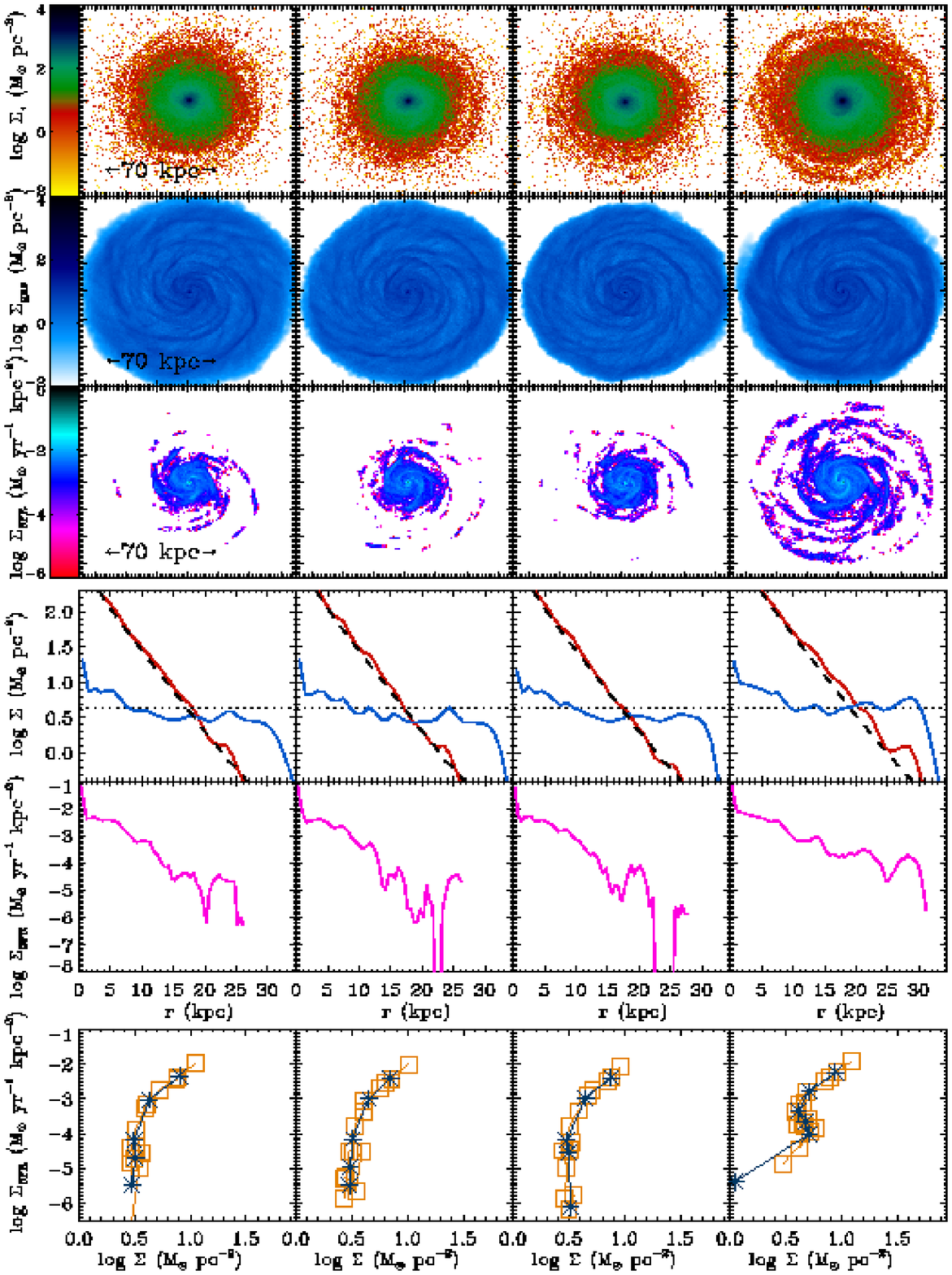}
\caption[]{ Maps of (from top to bottom) stellar mass surface
density, gas surface density, star formation rate surface density and plots of
stellar profile (red) and gas profile (blue), star formation rate
profile and Kennicutt-Schmidt law at 2.2 Gyr. The stellar mass profile from the
initial simulation is over-plotted on the mass profile plot as the
black dashed line. This is to highlight any differences from the
original profile. The simulations are (left to right)
exp-low with the disk mass equaling 0.041, 0.02
and 0.01 of the total mass, and exp-mid with a
disk mass equaling 0.041 of the total mass.  }
\end{figure}
\begin{figure}
\plotone{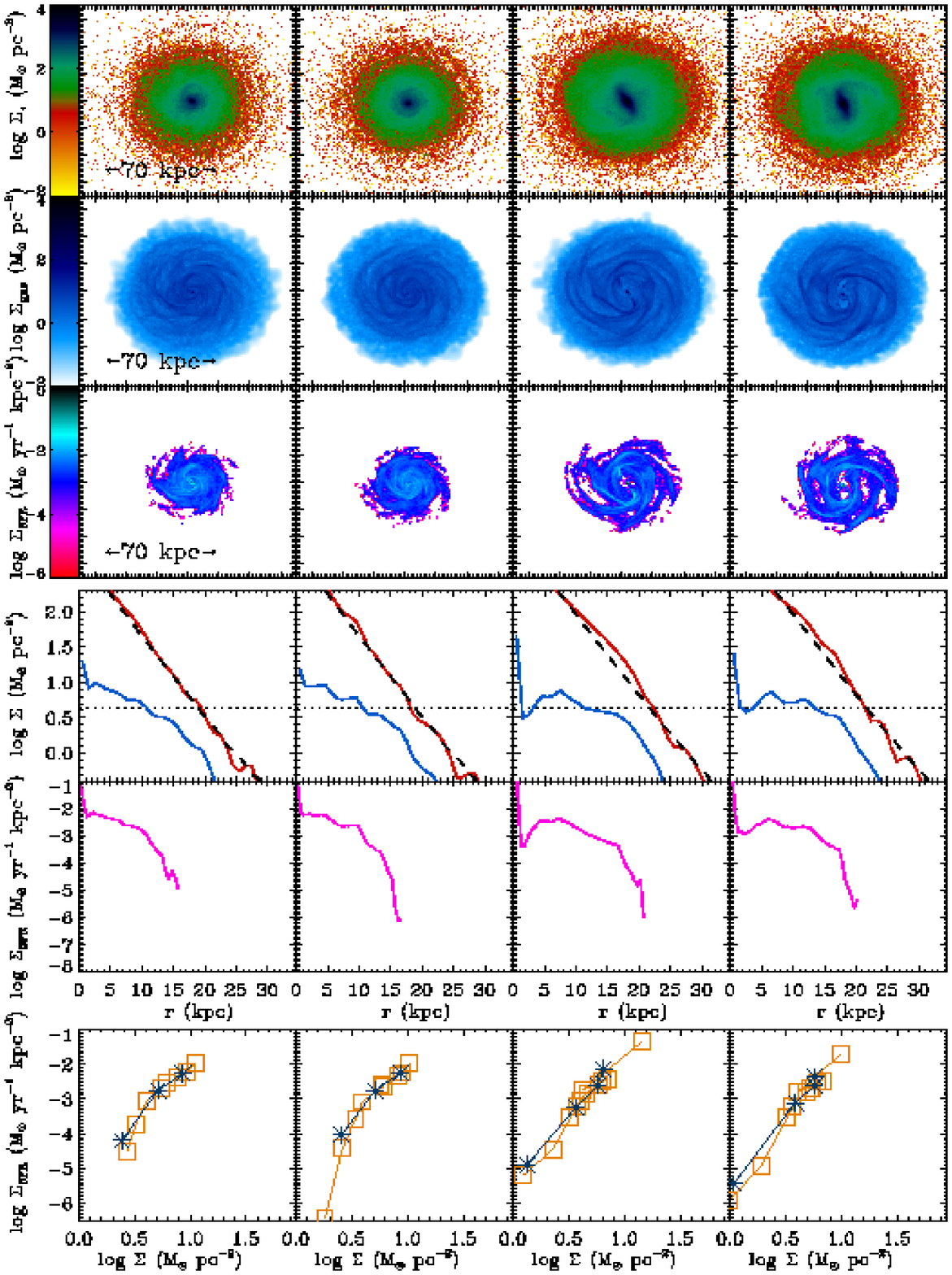}
\caption[]{Maps of (from top to bottom) stellar mass surface
density, gas surface density, star formation rate surface density and plots of
stellar profile (red) and gas profile (blue), star formation rate
profile and Kennicutt-Schmidt law at 2.2 Gyr. The stellar mass profile from the
initial simulation is over-plotted on the mass profile plot as the
black dashed line. This is to highlight any differences from the
original profile. The simulations are (left to right)
exp-mid with the disk mass equaling 0.02 and
0.01 of the total mass, and exp-high with a
disk mass equaling 0.041 and 0.02 of the total mass.  }
\end{figure}
\begin{figure}
\plotone{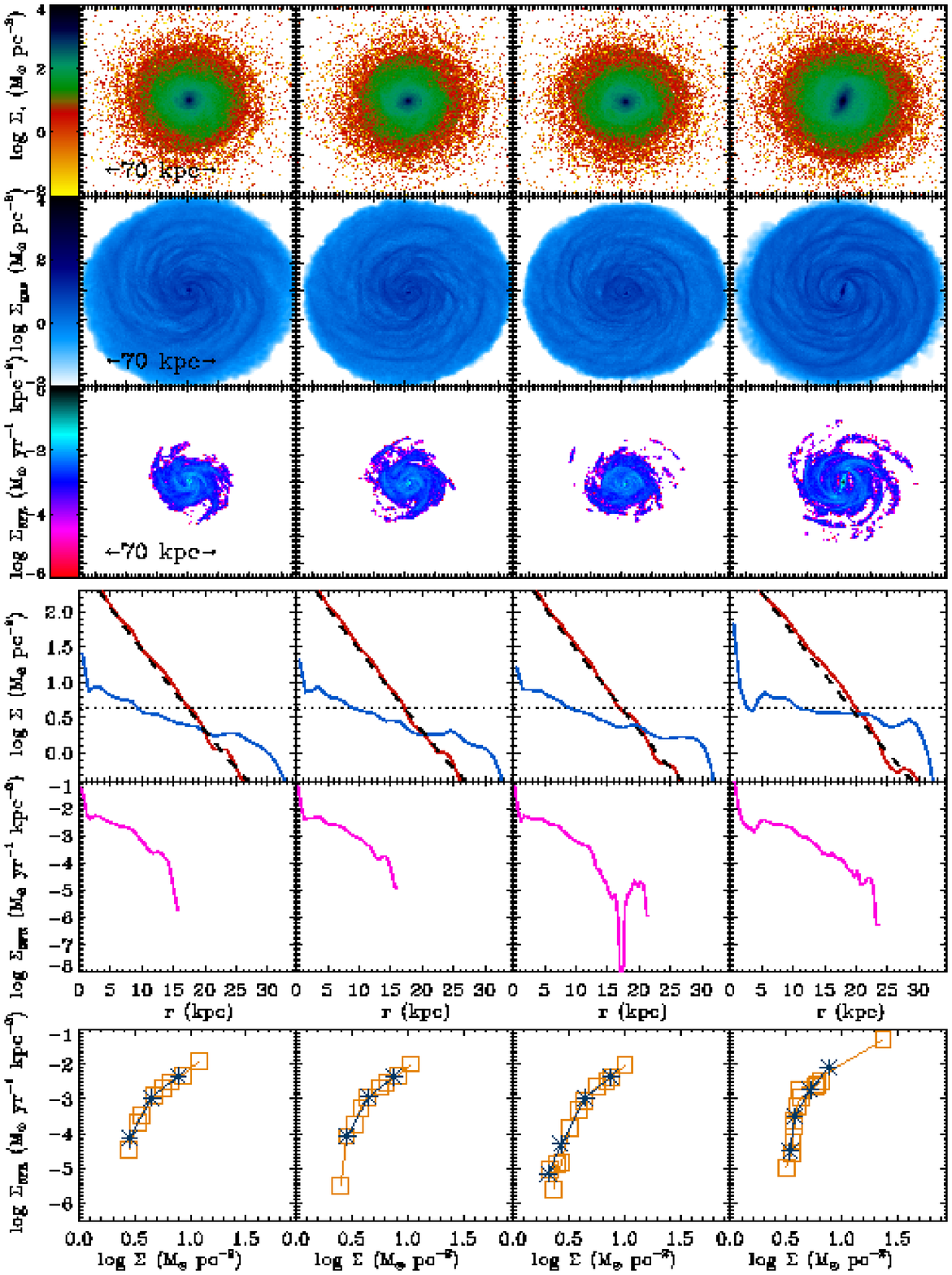}
\caption[]{Maps of (from top to bottom) stellar mass surface
density, gas surface density, star formation rate surface density and plots of
stellar profile (red) and gas profile (blue), star formation rate
profile and Kennicutt-Schmidt law at 2.2 Gyr. The stellar mass profile from the
initial simulation is over-plotted on the mass profile plot as the
black dashed line. This is to highlight any differences from the
original profile.  The simulations are (left to right) Mestel-low with the disk mass equaling 0.041, 0.02 and
0.01 of the total mass, and Mestel-mid with a disk
mass equaling 0.041 of the total mass.  }
\end{figure}
\begin{figure}
\plotone{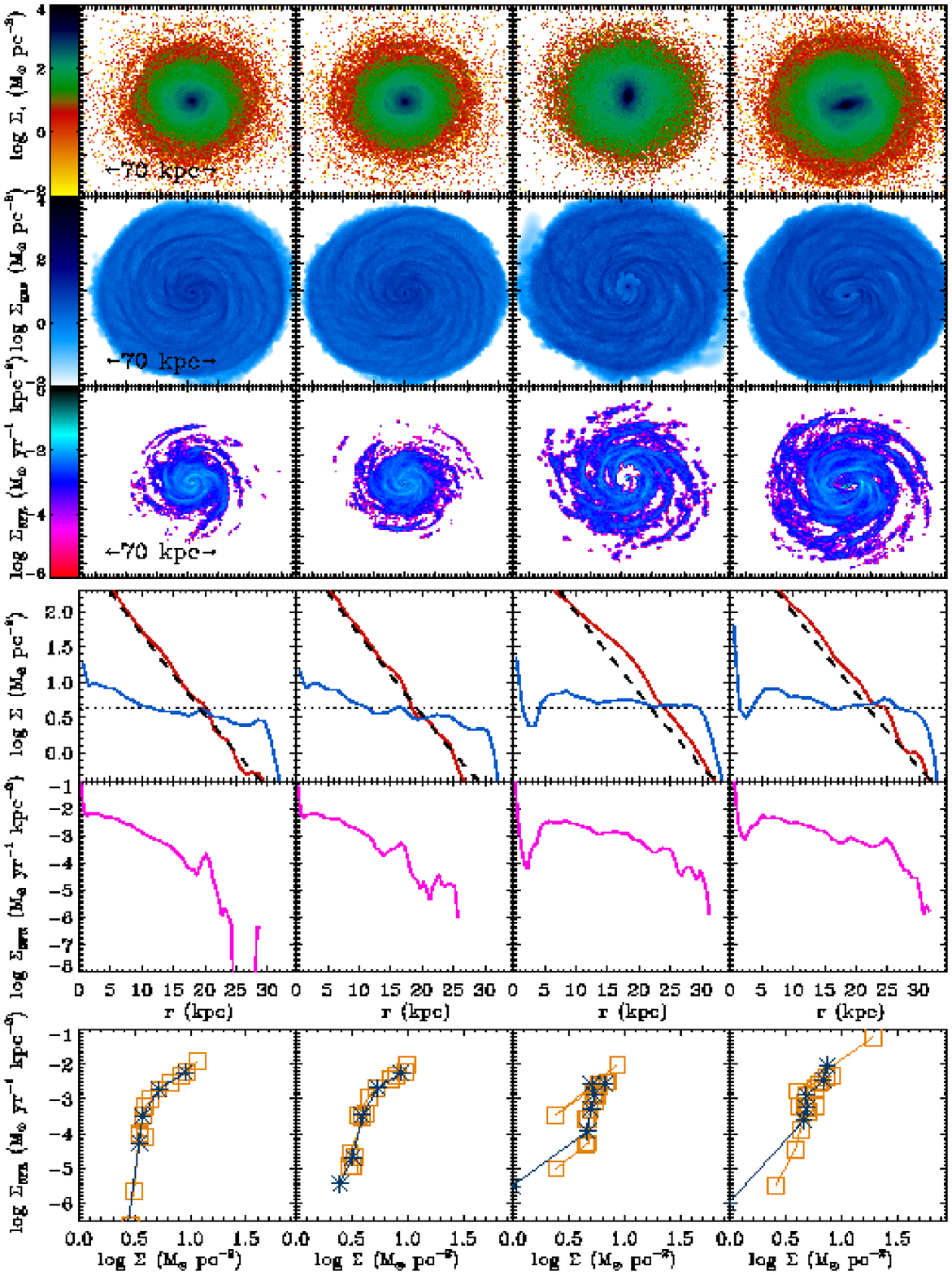}
\caption[]{Maps of (from top to bottom) stellar mass surface
density, gas surface density, star formation rate surface density and plots of
stellar profile (red) and gas profile (blue), star formation rate
profile and Kennicutt-Schmidt law at 2.2 Gyr. The stellar mass profile from the
initial simulation is over-plotted on the mass profile plot as the
black dashed line. This is to highlight any differences from the
original profile.   The simulations are (left to right)
Mestel-mid with the disk mass equaling 0.02 and 0.01 of
the total mass, and Mestel-high with a disk mass
equaling 0.041 and 0.02 of the total mass.  }
\end{figure}
\begin{figure}
\plotone{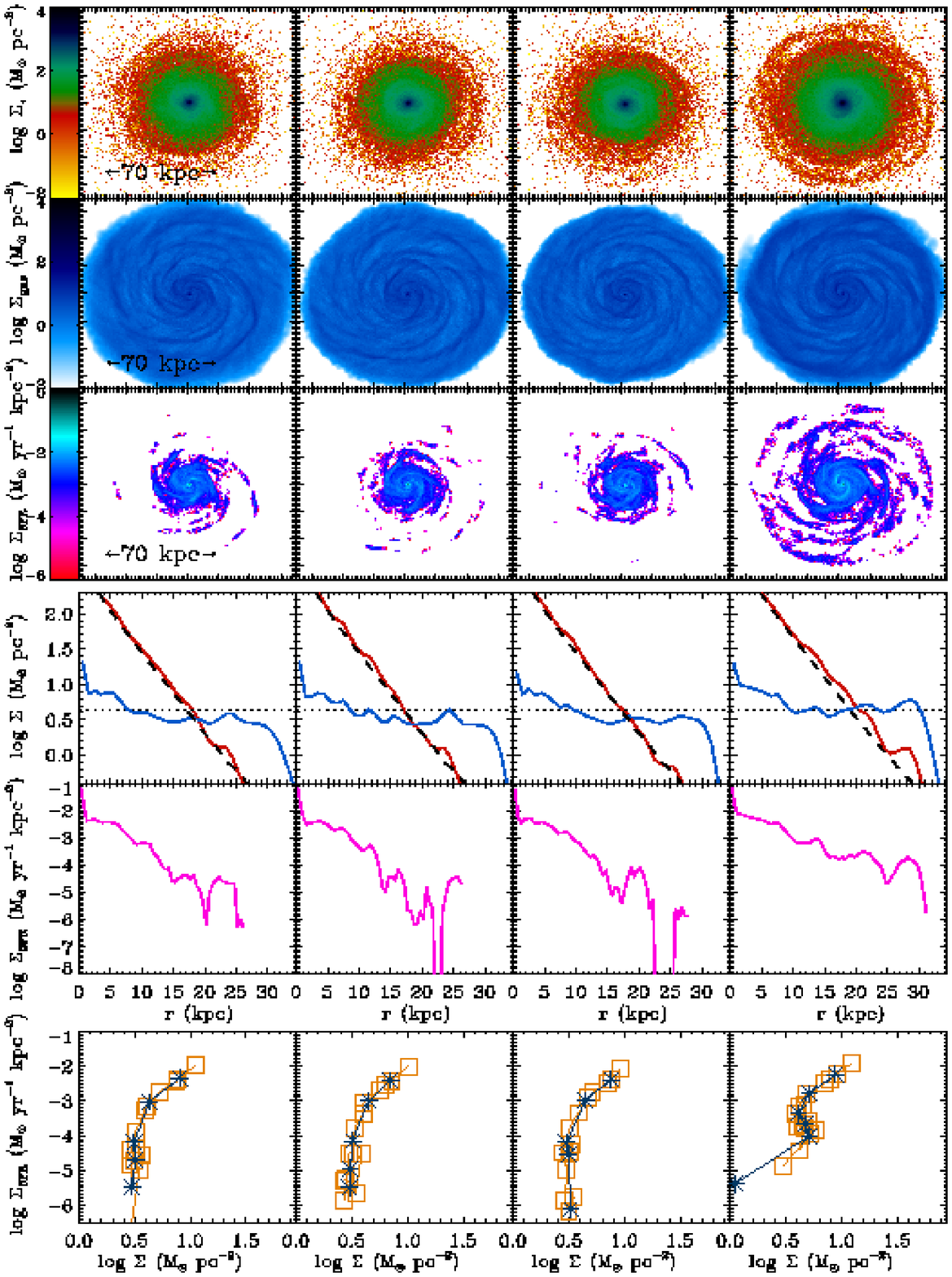}
\caption[]{ Maps of (from top to bottom) stellar mass surface
density, gas surface density, star formation rate surface density and plots of
stellar profile (red) and gas profile (blue), star formation rate
profile and Kennicutt-Schmidt law at 2.2 Gyr. The stellar mass profile from the
initial simulation is over-plotted on the mass profile plot as the
black dashed line. This is to highlight any differences from the
original profile.  The simulations are (Left to right)
flat-low with the disk mass equaling
0.041, 0.02 and 0.01 of the total mass, and flat-mid with a disk mass equaling 0.041 of the total mass.  }
\end{figure}
\begin{figure}
\plotone{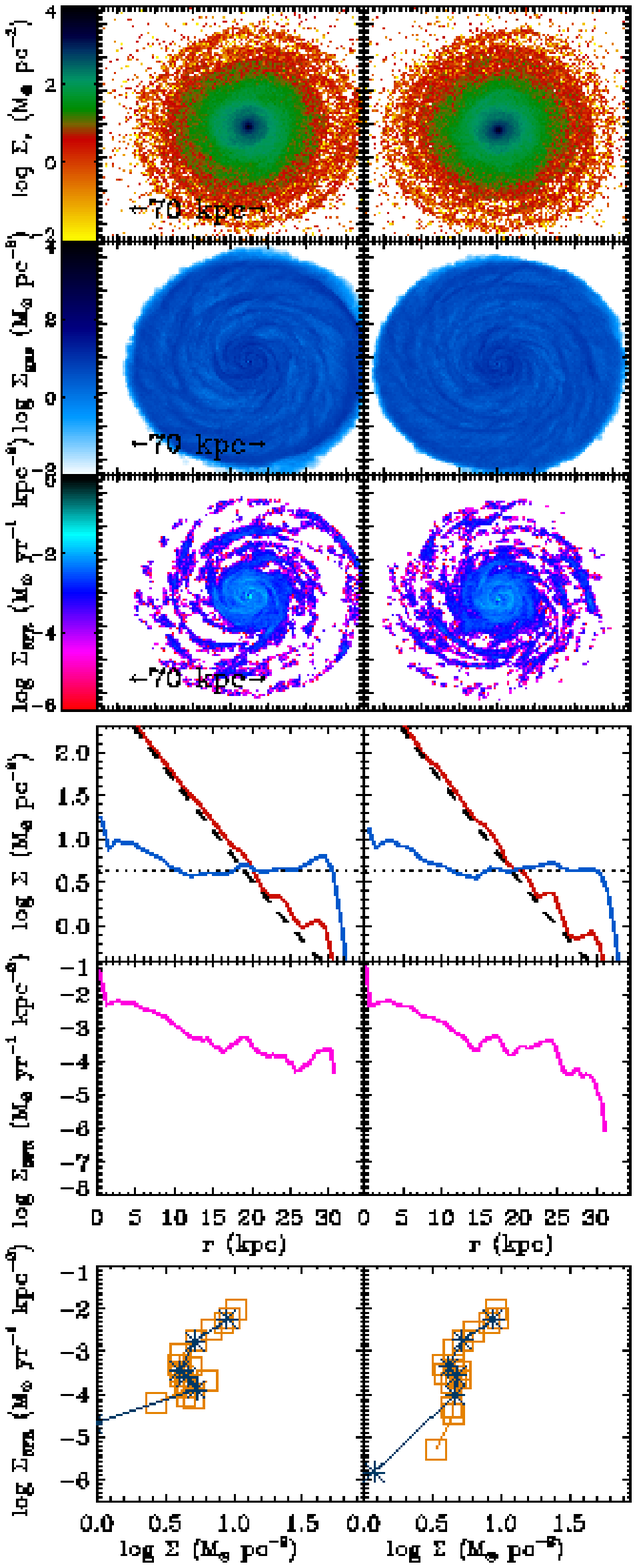}
\caption[]{Maps of (from top to bottom) stellar mass surface
density, gas surface density, star formation rate surface density and plots of
stellar profile (red) and gas profile (blue), star formation rate
profile and Kennicutt-Schmidt law at 2.2 Gyr. The stellar mass profile from the
initial simulation is over-plotted on the mass profile plot as the
black dashed line. This is to highlight any differences from the
original profile.  The simulations are (left to right)
flat-mid disk with the disk mass equaling 0.02
and 0.01 of the total mass.  }
\end{figure}

\end{document}